\documentclass[11pt]{article}
\usepackage{amssymb,amsthm,amsmath,float,subfigure}
\usepackage{appendix}
\usepackage{booktabs} 

\usepackage[total={6.5in,8.75in}, top=1.2in, left=0.9in, includefoot]{geometry}
\usepackage{graphicx}
\newcommand{\Eq}[1]{(\ref{eq:#1})}
\newcommand{\Th}[1]{Th.~\ref{thm:#1}}
\newcommand{\Lem}[1]{Lem.~\ref{lem:#1}}

\newcommand{\Sec}[1]{\S \ref{sec:#1}}
\newcommand{\Fig}[1]{Fig.~\ref{fig:#1}}
\newcommand{\Tbl}[1]{Table~\ref{tbl:#1}}
\newcommand{\App}[1]{App.~\ref{app:#1}}

\newcommand{\InsertFig}[4]
{\begin{figure}[h!t]
 \centerline{
 \includegraphics[width=#4]{./figures/#1}
 }
 \caption{{\footnotesize #2}
 \label{fig:#3}}
\end{figure}}

\newcommand{\InsertFigTwo}[5] {
\begin{figure}[h!t]
 \centerline{
 \includegraphics[width=#5]{./figures/#1}
 \hskip 0.1in
 \includegraphics[width=#5]{./figures/#2}
 }
 \caption{{\footnotesize #3}
 \label{fig:#4}}
\end{figure}}

\newcommand{\bC}{{\mathbb{ C}}}
\newcommand{\bN}{{\mathbb{ N}}}

\newcommand{\bQ}{{\mathbb{ Q}}}
\newcommand{\bR}{{\mathbb{ R}}}

\newcommand{\bT}{{\mathbb{ T}}}
\newcommand{\bZ}{{\mathbb{ Z}}}

\newcommand{\cD}{{\cal D}}

\newcommand{\cO}{{\cal O}}

\newcommand{\cT}{{\cal T}}

\newcommand{\eps}{\varepsilon}

\newcommand{\tr}{\mathop{\rm tr}}
\newcommand{\floor}[1]{{\lfloor{#1}\rfloor}}

\newcommand{\fix}[1]{\mathop{\mathrm{Fix}(#1)}}
\newcommand{\nint}{\mathop{\mathrm{nint}}}

\renewcommand{\vector}[3] {{\begin{pmatrix} {#1}\\{#2}\\{#3}\end{pmatrix}}}

\newtheorem{thm}{Theorem}
\newtheorem{lem}[thm]{Lemma}

\newcommand{\beq}[1]{\begin{equation}\label{eq:#1}}
\newcommand{\eeq}{\end{equation}}

\newenvironment{se}[1]{\equation\label{eq:#1}\aligned}{\endaligned\endequation}
\newcommand{\bsplit}[1]{\begin{se}{#1}}
\newcommand{\esplit}{\end{se}}

\newenvironment{example}[1][]
 {
	\setlength \leftmargini {1.0em}		
	\setlength \topsep {0.5em}			
	\begin{quote}
	{\it Example#1} }
	{\end{quote}
 }
\newcommand{\bexam}[1][:]{\begin{example}[#1]}
\newcommand{\eexam}{\end{example}}

\title{Greene's Residue Criterion for the Breakup of Invariant Tori of Volume-Preserving Maps}
\author{Adam M.~Fox\footnote{Adam.Fox@colorado.edu}
  \ and\ 
  James D.~Meiss\footnote{James.Meiss@colorado.edu}
	\ \thanks
 {
 AMF and JDM and were supported in part by NSF grant DMS-0707659. 
 AMF was also supported by the NSF MCTP grant DMS-0602284.
 Useful conversations with Hector Lomel\'i and Holger Dullin
 are gratefully acknowledged. 
 }
 \\
 	Department of Applied Mathematics\\
 University of Colorado \\
	Boulder, CO 80309-0526 \\
}
\date{\today}
\begin{document}
\maketitle

\begin{abstract}
\vspace*{1ex}
\noindent

Invariant tori play a fundamental role in the dynamics of symplectic and volume-preserving maps. Codimension-one tori are particularly important as they form barriers to transport. Such tori foliate the phase space of integrable, volume-preserving maps with one action and $d$ angles. For the area-preserving case, Greene's residue criterion is often used to predict the destruction of tori from the properties of nearby periodic orbits. Even though KAM theory applies to the three-dimensional case, the robustness of tori in such systems is still poorly understood. We study a three-dimensional, reversible, volume-preserving analogue of Chirikov's standard map with one action and two angles. We investigate the preservation and destruction of tori under perturbation by computing the ``residue" of nearby periodic orbits. We find tori with Diophantine rotation vectors in the ``spiral mean" cubic algebraic field. The residue is used to generate the critical function of the map and find a candidate for the most robust torus.
\end{abstract}

\section{Introduction}\label{sec:Introduction}
This paper is concerned with the breakup of invariant tori for volume-preserving maps. Tori are prominent as invariant sets for many dynamical systems. In particular a dynamical system is deemed to be ``integrable" when almost all of its orbits lie on tori and the dynamics on each torus is conjugate to a rigid rotation with some rotation vector $\omega$. This is the Arnold-Liouville formulation of integrability \cite{Arnold78}, and as such holds for Hamiltonian flows and symplectic maps, but also is essentially the notion of ``broad integrability" introduced by Bogoyavlenskij for more general systems \cite{Bogoy98}.

Here we will consider families of maps $f_\eps$ on $M = \bT^d \times \bR^k$ of the form
\bsplit{NearIntegrable}
	x' &= x + \Omega(z) - \eps h(x,z,\eps) ,\\
	z' &= z - \eps g(x,z,\eps) ,
\esplit
that have an ``integrable limit" when $\eps =0$.
Taking $\bT^d = \bR^d / \bZ^d$, the $x$ coordinates represent $d$-angles with unit period, and the $z$-coordinates represent $k$ action-like variables. Thus the map \Eq{NearIntegrable} is said to be a $d$-angle, $k$-action map.
We will assume that $f$ is volume-preserving, i.e., that its Jacobian satisfies
\[
	\det{Df_\eps} = 1. 
\]
Examples of such maps include Chirikov's standard map (with $d=k=1$) 
\bsplit{StdMap}
	x' &= x+z-\eps \sin(2\pi x) ,\\
	z' &= z - \eps \sin(2\pi x) .
\esplit
This map is area-preserving, and is thus both volume-preserving and symplectic.
Another oft-studied case is the four-dimensional symplectic map due to Froeshl\'e (with $d=k=2$) \cite{Froeschle73}. The dynamics for $d+k=3$ applies to the motion of a passive scalar in an incompressible fluid \cite{Piro88,Cartwright96}, as well as granular mixing \cite{Meier07} and magnetic field-line flows \cite{Thyagaraja85}. 
In this paper we will study a two-angle, one-action map---a $(2+1)$-dimensional map---that models the typical dynamics near a rank-one resonance \cite{Dullin12}. Our model is introduced in \Sec{Model}.

The \emph{frequency map}
\beq{FrequencyMap}
	\Omega: \bR^k \to \bR^d
\eeq
plays a key role in the dynamics of $f_\eps$. In particular every orbit of the integrable limit, $f_0$, lies on a $d$-dimensional torus,
$\cT_z = \bT^d \times \{ z \}$, on which the dynamics is simply a rigid translation, $x_t = x_0 + \omega t$ with $\omega = \Omega(z_0)$. We call any $d$-torus that is homotopic to $\cT_0$, \emph{rotational}.
When the rotation vector $\omega$ is \emph{incommensurate}, namely
\[
	\{(p,q) \in \bZ^d \times \bZ: p \cdot \omega = q \} = \{(0,0)\} ,
\]
such orbits are dense on $\cT_z$. KAM theory studies the persistence of the rotational tori of perturbations of integrable system like $f_0$ \cite{Broer90, Poshel01, delaLlave01}. The application of this theory requires that $f_\eps$ preserve some structure (e.g., is symplectic or reversible), is smooth enough, has a nondegenerate frequency map, and finally that the rotation vector of the tori be sufficiently irrational, e.g., $\omega \in \cD_{s}$ is Diophantine.
\beq{Diophantine}
	\cD_{s} = \left\{\omega \in \bR^d: \exists\, c > 0 \;\;\mbox{s.t. } |p\cdot \omega-q| > \frac{c}{|p|^s}\right\},
\eeq
for some $s \ge d$.
Since we assume that $k=1$, whenever a rotational invariant torus exists it has codimension-one and therefore is a barrier to transport. 

The orbits of $f_0$ with rational rotation vectors, $\omega \in \bQ^d$, are periodic.
More generally, each periodic orbit of \Eq{NearIntegrable} can be assigned a rotation vector by lifting the map to $\bR^d \times \bR^k$. The lift of a period-$n$ orbit obeys
\beq{periodicLift}
	(x_n,z_n) = (x_0+m,z_0),
\eeq
for $m \in \bZ^d$, and thus has rotation vector $\omega = m/n$. We call these $(m,n)$-periodic orbits, see \Sec{PeriodicOrbits}.

Following John Greene \cite{Greene79}, we study the persistence of rotational tori by finding a sequence of rational vectors $m_\ell/n_\ell$ that limit on a given incommensurate vector $\omega$. Greene hypothesized that the limit of the corresponding $(m_\ell,n_\ell)$-periodic orbits of $f_\eps$ is a quasiperiodic orbit dense on an invariant torus if and only if the periodic orbits remain stable in the limit; more specifically, if their \emph{residues} remain bounded. This is \emph{Greene's residue criterion}, see \Sec{ResidueCriterion}. 
As we recall in that section, proofs of this result are known in certain situations. 

Moreover, Tompaidis generalized Greene's criterion to higher-dimensional symplectic and quasi\-periodically-forced symplectic maps \cite{Tompaidis96a}. Indeed, the residue criterion has been applied to study the breakup of two-tori for several four-dimensional models, e.g., the Froeshl\'e map \cite{Bollt93, Borgonovi93, Tompaidis96b, Kurosaki97, Tompaidis99, Zhou01b, Zhou01a, Celletti04}, and a quadratic map \cite{Vrahatis97}. We are not aware of previous use of the residue criterion for volume-preserving maps apart from a quasi\-periodically-forced, area-preserving map \cite{Artuso91, Artuso92, Tompaidis96b}, a $(2+1)$-dimensional map with one of the components of $\Omega$ set to a fixed irrational value.

As we discuss in \Sec{Reversible}, our model---like the standard map \Eq{StdMap}---is reversible, and this makes finding periodic orbits especially easy: a two-dimensional secant method suffices, see \Sec{Numerics}. In addition, since one of the multipliers is always $1$ for symmetric orbits, there is a natural definition of the residue, see \Sec{VPResidue}. 

One complicating feature of \Eq{NearIntegrable} with $k < d$ is that the image of the frequency map \Eq{FrequencyMap} is at most a $k$-dimensional subset of the $d$-dimensional space of rotation vectors; a consequence is that orbits with given rotation vectors typically do not exist for fixed parameters. In our $(2+1)$-dimensional model, we get around this by adding a new parameter, $\delta$, to $\Omega$ so that $\Omega: (z,\delta) \mapsto (\omega_1,\omega_2)$ is a diffeomorphism. With this modification, we observe, in \Sec{Numerics}, that there are symmetric periodic orbits for any $\eps$ and any $m/n \in \bQ^2$.

We will show in \Sec{NumericalResidues} that the residues of high period orbits appear to undergo a rapid transition from ``nearly zero" to ``exponentially large" as $\eps$ grows, just like for the standard map. In \Sec{FareySequences} we use this transition to get a reasonably sharp estimate for a critical set of parameters at which a given torus is apparently destroyed. 

The tori that we study in \Sec{FareySequences} have rotation vectors that are integral bases for a cubic algebraic field $\bQ(\sigma)$, where $\sigma$ is the ``spiral mean" \cite{Kim86}, see \Sec{SpiralMean}. It has long been conjectured that simple cubic irrationals like the spiral mean could be the analogue of the golden mean for the standard map \Eq{StdMap}. Indeed, most of the previous studies of Greene's criteria for multidimensional tori have used vectors in $\bQ(\sigma)$, though there have been several that studied other cubics \cite{Tompaidis96b, Zhou01b,Zhou01a} and even quartic irrationals \cite{Vrahatis97, Kurosaki97}.

Another remarkable conjecture in \cite{Greene79} is that the last invariant circle of \Eq{StdMap} has the golden mean rotation number. There is strong numerical support for this conjecture, and more generally for the conjecture that circles with ``noble" rotation vectors appear to be locally most robust \cite{MacKay92c}. Do these notions have a higher-dimensional generalization? We are not aware of any previous progress on this question. In \Sec{LastTorus} we look for the last torus for our model by finding the set of critical parameter values, $\eps_{cr}(\omega)$, for a set of tori whose rotation numbers are integral bases of $\bQ(\sigma)$. We use Kim and Ostlund's generalization of the Farey tree, see \Sec{KimOstlund}, to systematically generate sets of Diophantine rotation vectors.

\section{Greene's Residue Criterion}\label{sec:ResidueCriterion}

John Greene studied the persistence and destruction of rotational invariant circles of the standard map \Eq{StdMap} \cite{Greene68, Greene79} by approximating them with sequences of periodic orbits. The Poincar\'e-Birkhoff theorem implies that the standard map has at least two $(m,n)$-periodic orbits for any choice of rational rotation number $\tfrac{m}{n}$ \cite{Meiss92}. These orbits, which exist for all $\eps$, are often called the \emph{Birkhoff orbits}. Since the standard map is area-preserving, the product of the multipliers of any period-$n$ orbit is $1$, and so its stability can be completely characterized by a quantity Greene called the residue:
\beq{GreenesResidue}
	R = \tfrac14( 2-\tau), \; \mbox{where } \tau = \tr(Df^n(x,z)).
\eeq
The residue conveniently encodes the stability of an orbit: it is elliptic (complex, unit modulus multipliers) when $0 < R < 1$, hyperbolic (real, positive multipliers) when $R < 0$, and reflection hyperbolic (real, negative multipliers) when $R > 1$. One of the two Birkhoff orbits of \Eq{StdMap} has positive residue while the other has negative residue, and Greene showed that for a period-$n$ orbit of \Eq{StdMap}, 
\beq{AsymptoticResidue}
	R = \cO(\eps^n),
\eeq
both for $\eps \ll 1$ and for $\eps \gg 1$. 

Greene's $6^{th}$ assertion, now known as \emph{Greene's Residue Criterion}, is perhaps the most astonishing of the conjectures in \cite{Greene79}. In particular, consider an irrational rotation number $\omega \in \bR$, with continued fraction $\omega = [k_0,k_1,\ldots] = k_0 + 1/(k_1 + \ldots)$, and let
\beq{Convergents}
	\frac{m_\ell}{n_\ell} = [k_0,k_1,\ldots,k_l]
\eeq
be its $\ell^{th}$ convergent. Thus the $(m_\ell,n_\ell)$-Birkhoff orbits, with residues $R^\pm_\ell$, have rotation numbers that converge to $\omega$. One formulation of Greene's criterion is:

\begin{quote}\textbf{Greene's Residue Criterion}: A rotational invariant circle of an area-preserving twist map with a given irrational rotation number $\omega$ exists if and only if the residues of its convergent Birkhoff orbits, $R^\pm_\ell$, remain bounded as $\frac{m_\ell}{n_\ell} \to \omega$.
\end{quote}
\noindent
A stronger version of this criterion asserts that when there is an invariant circle the mean residue,
\beq{meanResidue}
	\mu(\omega) = \lim_{\ell \to \infty} \frac{1}{n_\ell} \log|R_\ell| ,
\eeq
exists and is negative for any sequence of $(m_\ell,n_\ell)$-orbits whose rotation number converges to $\omega$.

Greene studied in particular the invariant circle with golden mean rotation number
\beq{GoldenMean}
	\phi = \tfrac12(1+\sqrt{5})= [1,1,1,\ldots].
\eeq
For the convergents to $\phi$, numerical studies show that there is a parameter value $\eps_{cr}(\phi)$ such that $R^\pm_\ell \to 0$ as $\ell \to \infty$ whenever $\eps < \eps_{cr}(\phi)$. Conversely, whenever $\eps > \eps_{cr}(\phi)$ the residues grow exponentially with the period, and there appears to be no golden invariant circle. Greene estimated $\eps_{cr}$ by computing a sequence of parameter values for which $|R_l^\pm|$ reaches some fixed value $R_{th} > 0$, obtaining
\beq{kCritical}
	\eps_{cr}(\phi) \approx \frac{1}{2\pi} 0.97163540631.
\eeq	
The value of the threshold $R_{th}$ is irrelevant, but Greene found that $R_{th} \approx 0.25$ gave the most rapid convergence.

Some aspects of Greene's residue conjecture have been proven.

\begin{thm}[Residue Criterion for Twist Maps \cite{MacKay92a, Falcolini92}]\label{thm:Residue}
Suppose $f_\eps$ is an analytic, area-preserving twist map, and the sequence of $(m_\ell,n_\ell)$-orbits converges to an analytic invariant circle on which the dynamics is analytically conjugate to rigid rotation with rotation number $\omega \in \cD_s$, then 
there are constants $C,K > 0$ such that
\[
	|R_l| < C \exp \left(-K |\omega-m_\ell/n_\ell|^{-1/(1+s)}\right)
\]
\end{thm}

\noindent
In particular since $|\omega -m_\ell/n_\ell| < c/n_\ell^2$ for continued fraction convergents, this implies that $R_l \to 0$ exponentially in the period. The residue criterion also applies to twist-reversing 
maps \cite{Delshams00}, as studied numerically by \cite{delCastillo96,Apte03}.

Aspects of the converse statement of the residue criterion have also been proven for area-preserving twist maps. From Aubry-Mather theory, these maps have a set of ``minimizing" orbits for each $\omega$, and when $\omega$ is irrational this set is either a circle or a Cantor set---a cantorus. If the cantorus has a positive Lyapunov exponent, then there exists a sequence $(m_\ell,n_\ell)$ such that the mean residue \Eq{meanResidue} is positive and has the value of this exponent \cite{Falcolini92}. Moreover, if the cantorus is uniformly hyperbolic, then this sequence can be taken to be the sequence of minimizing orbits (the negative residue Birkhoff orbits) \cite{MacKay92a}. 

The residue criterion has also been generalized to higher dimensions.
For example, a $2d$ dimensional symplectic map has \emph{partial residues}
\[
	R^{(j)} = \tfrac14(2-\lambda_j - \lambda^{-1}_j) \;, \quad j = 1,\ldots, d
\]
for each of the reciprocal pairs of multipliers of a given orbit \cite{Kook89}.
Tompaidis \cite{Tompaidis96a} proved that if $f_\eps$ is a $C^r$ symplectic map with $r>1$, and the twist, $D\Omega$, is nondegenerate then, when there is a $C^r$, Diophantine, invariant $d$-torus, the partial residues of any $(m,n)$-periodic orbits with rotation numbers sufficiently close to $\omega$ obey the bound
\[
	|R^{(j)}| < C n |n\omega - m|^k
\]
for any positive integer $k < (r-1)/2s$. His results also apply to the rotating standard map of \cite{Artuso92}. 

We are not aware of any theorem, however, that applies to volume-preserving maps more generally. Thus we turn to numerical investigations of a model.

\section{One-Action Maps: A Three-Dimensional Model}\label{sec:Model}

We will study the rotational tori of a $(2+1)$-dimensional map of the form \Eq{NearIntegrable}.
When $k = 1$ the image of a frequency map $\Omega$ for \Eq{NearIntegrable} is a curve. Near resonance, $p \cdot \Omega(z) = q$, the rotational invariant tori of \Eq{NearIntegrable} are typically fragile.
The resulting local dynamics is strongly influenced by whether the curve $\Omega(z)$ is transverse to or tangent to such a resonance \cite{Dullin12}.
We will study dynamics of the two-angle case with the frequency map
\beq{Omega}
	\Omega(z,\delta) = (z + \gamma, \beta z^2 -\delta),
\eeq
which, as was shown in \cite{Dullin12}, is a normal form for dynamics near a ``rank-one" resonance. When $\beta \neq 0$, the map \Eq{Omega} satisfies the nondegeneracy condition 
\beq{Nondegeneracy}
	\det (D_z\Omega, D_z^2\Omega, \ldots, D_z^d\Omega) \neq 0
\eeq
that is sufficient for KAM theory \cite{Cheng90a, Xia92}.
For fixed $(\gamma, \beta, \delta)$, the image $\Omega: z \mapsto \omega$ is a parabola. However, when thought of as a map $\Omega: (z,\delta) \mapsto (\omega_1,\omega_2)$, $\Omega$ becomes bijective. The parameter $\delta$ is also important in another sense: it unfolds a tangency with the $(0,1,0)$ resonance (at $z=\delta = 0$). As was shown in \cite{Dullin12}, near a resonant tangency, even though the ``twist" condition \Eq{Nondegeneracy} applies, the dynamics is like that of nontwist maps.
Consequently, we will take $\delta$ to be an essential parameter, but will fix $\beta$ and $\gamma$. 

 Another requirement for the preservation of the rotational tori of \Eq{NearIntegrable} for nonzero $\eps$ is that the map satisfy an ``intersection property". A necessary condition is that the force has zero average,
\beq{Exact}
	\int_{\bT^d} g(x,z,\eps) dx = 0 .
\eeq
Indeed, if \Eq{Exact} is not satisfied, then $f_\eps$ may have no invariant tori; for example, when $g(x,z,\eps) = const \neq 0$, then the $z$ coordinates of \Eq{NearIntegrable} drift and there are no recurrent orbits for any $\eps \neq 0$.
For the two-angle case, we will use the simple form for the force, 
\beq{Force}
	g(x) = a \sin(2\pi x_1) + b \sin(2\pi x_2) + c \sin( 2\pi (x_1 - x_2)) .
\eeq
The three terms in $g$ represent resonant forcing for $(p,q) = (1,0,q)$, $(0,1,q)$ and $(1,-1,q)$, respectively, for each $q \in \bZ$. The last term explicitly couples the two angles $(x_1, x_2)$; however, note that even when $c = 0$ these are coupled through the frequency map \Eq{Omega}.

As a final simplification, we will choose the perturbation $h$ in \Eq{NearIntegrable} so that $\eps h \equiv \Omega(z)-\Omega(z')$, giving the model of \cite{Dullin12}
\bsplit{VPMap}
	x' &= x+ \Omega(z',\delta) \mod 1 ,\\
	z' &= z - \eps g(x) .
\esplit
An advantage of this form is that it is always a homeomorphism; indeed, it has the inverse
\beq{VPInverse}
	f_\eps^{-1}(x,z) = (x-\Omega(z,\delta),z + \eps g(x-\Omega(z,\delta))) .
\eeq
In addition, this map is an exact-volume-preserving diffeomorphism whenever $\Omega$ and $g$ are $C^1$ and $g$ satisfies \Eq{Exact}.

One could also regard $f$ as a diffeomorphism on $\bT^2 \times \bR^2$, by adjoining the trivial dynamics $\delta' = \delta$ to \Eq{VPMap}. 

Following \cite{Meiss12a}, we will typically use the ``standard" set of parameters 
\beq{StdParams}
	a=b=c=1, \; \beta = 2, \; \gamma = \tfrac12(\sqrt{5}-1) ,
\eeq
and think of $\eps$ and $\delta$ as the parameters that govern the strength of the forcing and the unfolding of the tangency, respectively.

Unless otherwise mentioned, all the computations below are for \Eq{VPMap} with the frequency map \Eq{Omega}, the force \Eq{Force}, and the standard parameters \Eq{StdParams}.

%
%
%

\subsection{Periodic Orbits}\label{sec:PeriodicOrbits}
Rotational periodic orbits of \Eq{VPMap} are partially classified by their rotation vectors. Let $F$ be the lift of $f$ to the universal cover $\bR^3$ of $M$ obtained by simply removing the mod 1 from \Eq{VPMap}. A sequence
\beq{Orbit}
	\Gamma = \{(x_{t},z_{t}) = F(x_{t-1}, z_{t-1}): t \in \bZ\}
\eeq
is a type $(m,n) \in \bZ^2 \times \bN$ periodic orbit of $F$ if it obeys \Eq{periodicLift}.
Alternatively, noting that the rigid translation operator
\beq{Rigid}
	T_m(x,z) = (x + m,z) ,
\eeq
is a symmetry of $F$, then the periodicity condition becomes $T_{-m}F^n(x,z) = (x,z)$. 
If $m$ and $n$ are coprime, i.e., $gcd(m_1,m_2,n) = 1$, the projection of each such orbit onto $M$ is a period-$n$ orbit of $f$. 

The rotation vector of an orbit is the average increase in the angle per iteration,
\[
	\omega(\Gamma) = \lim_{t\to\infty} \frac{1}{t}( x_t - x_0)
\]
if this limit exists. Each $(m,n)$-orbit has rotation vector $\frac{m}{n}$.

An $(m,n)$-orbit of \Eq{VPMap} with the frequency map \Eq{Omega} must satisfy the three equations
\bsplit{POEqs}
	m &= \sum_{t=0}^{n-1} \Omega(z_t,\delta) \Rightarrow 
		\left\{ \begin{array}{l} m_1 = n \gamma + \sum_{t=0}^{n-1} z_t \\
								m_2 = -n\delta + \beta \sum_{t=0}^{n-1} z_t^2 \\
				\end{array} \right. , \\
	0 &= \sum_{t=0}^{n-1} g(x_t) .
\esplit
with $(x_t,z_t) = F^t(x_0,z_0)$. Ideally a system of three equations in three unknowns has isolated solutions; however, we recall again that for any fixed $\delta$ and a given $(m,n)$, there are typically no solutions of \Eq{POEqs}. For example, when $\eps = 0$, the curve $\Omega(z,\delta)$ will typically not intersect the point $\frac{m}{n}$, and there will be no $(m,n)$-orbit. However, including the parameter $\delta$ in \Eq{Omega}, there is an orbit for $\eps = 0$ at
\beq{ZDeltaStar}
	(z^*,\delta^*) = \left(\frac{m_1}{n} - \gamma, \beta {z^*}^2 -\frac{m_2}{n}\right),
\eeq
and any value for $x$. This corresponds to a two-torus of $(m,n)$-orbits of $f$. 

More generally, if an orbit satisfies the first and last equations in \Eq{POEqs}, then the value of $\delta$ is completely determined by the second of these equations, which can be rewritten as
\beq{DeltaEq}
	\delta = \frac{1}{n} \sum_{t=0}^{n-1} z_t^2 -\omega_2
		 \equiv \left< z^2 \right> - \omega_2.
\eeq
Thus $\delta$ is fixed by the mean square average of the action along the orbit. Of course the system \Eq{POEqs} cannot be solved independently of $\delta$, since the iterates of \Eq{VPMap} depend on its value. As we will see below, with $\delta$ included as a varying parameter, it appears that $f_\eps$ has orbits for all $(m,n) \in \bZ^2 \times \bN$.

For example, fixed points of \Eq{VPMap} occur with $(z^*,\delta^*)$ given by \Eq{ZDeltaStar} upon setting $n=1$. Thus these values are uniquely determined for any $(m_1, m_2)$. The angles are then determined by $g(x^*) = 0$, which generically has a one-dimensional set of solutions that project to circles on $\bT^2$. For example, if $a=b=c$ fixed points occur on the three circles $x_1 = 0$, $x_2 = \tfrac12$, and $x_1-x_2 = \tfrac12$. More generally there is at least a pair of circles of fixed points; an example is shown in \Fig{fixedPoints}. 

\InsertFig{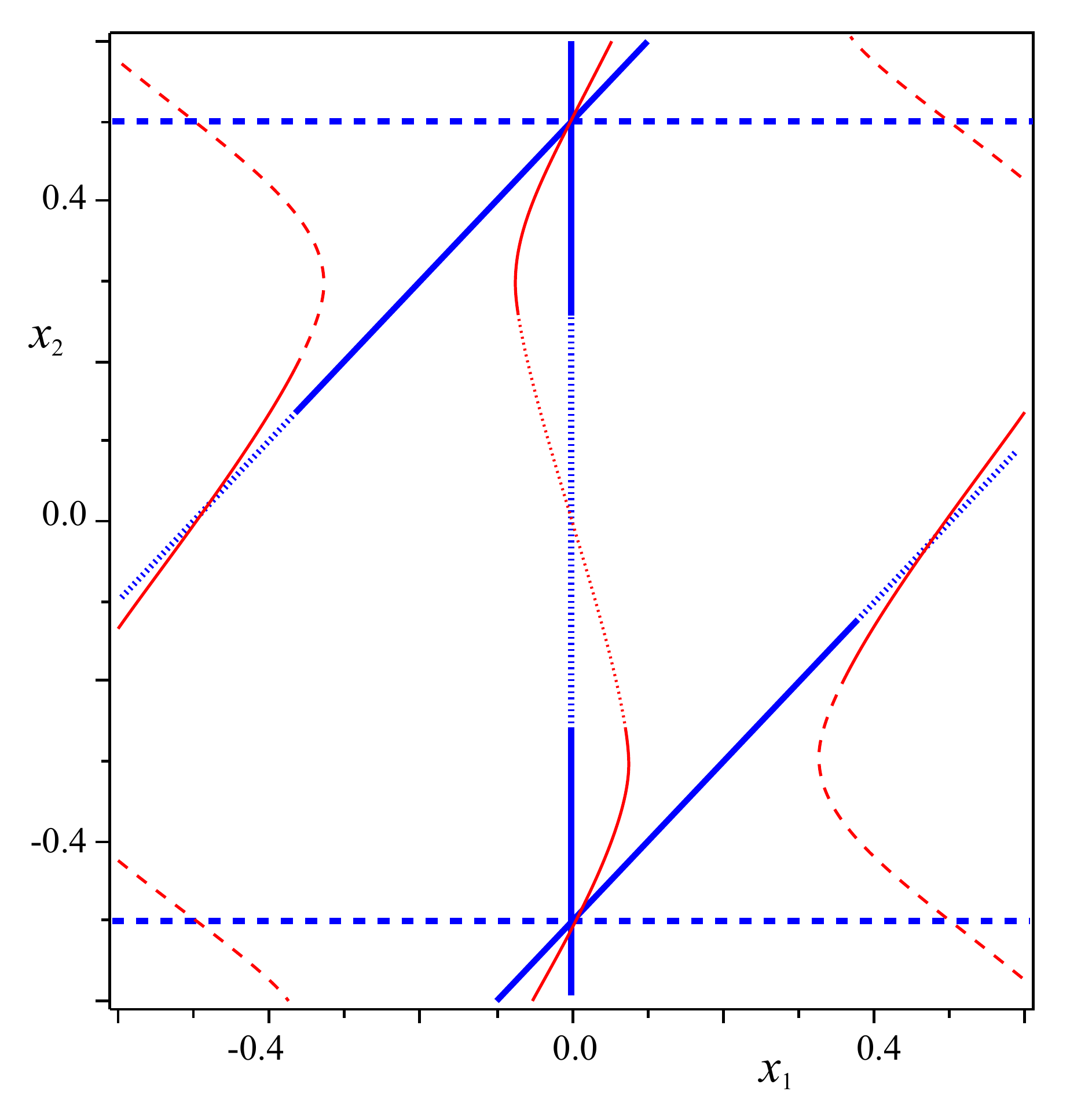}{Curves of fixed points for \Eq{VPMap} projected onto the angles $x$. The thick (blue) lines correspond to the case $a=b=c$, and the thinner (red) curves to $(a,b,c) = (1.0,0.7,0.3)$. The stability of these orbits is indicated for $\eps = 0.7$ and $m_1 = 1$; unstable orbits with $R < 0$ are shown as dashed lines and those with $R > 1$ as dotted lines. }{fixedPoints}{3.4in}

For period two, the system \Eq{POEqs} becomes
\bsplit{PeriodTwo}
	z_0 &= \tfrac12(m_1 + \eps g(x_0)) -\gamma,\\
	z_1 &= \tfrac12(m_1 - \eps g(x_0)) -\gamma, \\
	x_1 &= x_0 + \Omega(z_1,\delta), \\
	\delta &= \tfrac12(\beta z_0^2 + \beta z_1^2 - m_2)
\esplit
leaving one transcendental equation
\[
	g(x_0) + g(x_1) = 0
\]
for $x_0$. 
Consequently, there is again a set of curves of period-two orbits. Note however, that the the value of $\delta$ depends upon $\eps$ and $m$ for $n=2$. This holds more generally for higher period orbits as well.

\subsection{Reversibility}\label{sec:Reversible}
As for the standard map \Eq{StdMap}, it is very convenient that the map \Eq{VPMap} with \Eq{Force} is reversible as this simplifies the search for periodic orbits. Recall that a map is reversible if it is conjugate to its inverse, that is if there exists a homeomorphism $S$ such that $f \circ S = S \circ f^{-1}$ . The collection of the symmetries and reversors of a map form a group, the \emph{reversing symmetry} group \cite{Roberts92, Lamb98}.

Whenever $g$ is an odd function (as in \Eq{Force}), the map \Eq{VPMap} is reversible, with the two reversors
\bsplit{Involution}
	S_1(x,z) &= (-x,z - \eps g(x)) , \\
	S_2(x,z) &= (-x +\Omega(z,\delta), z) .
\esplit
To see this note that
\[
 S_2(x,z) = f \circ S_1(x,z)= S_1 \circ f^{-1}(x,z) ,
\]
since the inverse of $f$ is \Eq{VPInverse}. 
Since $g$ is odd, both $S_1$ and $S_2$ are involutions, that is $S_i^2 = id$, consequently 
\[
	f = S_2 \circ S_1 .
\]
Factorization into a pair of involutions also holds for Chirikov's standard map \Eq{StdMap}, and this was exploited by Greene and MacKay \cite{MacKay93}.

An orbit \Eq{Orbit} is \emph{symmetric} with respect to a reversor $S$ if $S(\Gamma) = \Gamma$; i.e., there exists a $j \in \bZ$ such that $S(x_0,z_0) = (x_j,z_j)$. Denoting the fixed sets of the reversors by $\fix{S_i}$, it is well-known that any symmetric orbit necessarily has points on these sets, see \App{appendix}. Indeed, for the lift, there are two points on two fixed sets that occur essentially halfway around the orbit (depending on whether $n$ is even or odd), and essentially halfway along in the angle direction (depending on whether the components of $m$ are even or odd), see \Tbl{Symmetry}. For the reversors \Eq{Involution} and translation symmetry \Eq{Rigid}, the relevant fixed sets are the curves
\bsplit{FixedSets}
	\fix{S_1 \circ T_{-m}} &= \{(\tfrac{m}{2},z): z \in \bR\} , \\
	\fix{S_2 \circ T_{-m}} &= \{(\tfrac12(\Omega(z,\delta)+m), z): z \in \bR) .
\esplit
Projecting back to the torus $\bT^2$, we see there are eight symmetry curves---four for $S_1$, the lines above $x = (0,0)$, $(\tfrac12,0)$, $(0,\tfrac12)$ and $(\tfrac12,\tfrac12)$, and four curves for $S_2$. Since symmetric orbits intersect these lines in pairs, as shown in \Tbl{Symmetry}, we expect to find four distinct symmetric periodic orbits for each $(m,n)$.

Moreover, since the one-dimensional fixed sets are graphs over the action coordinate, $z$, any rotational invariant torus of $f$ necessarily intersects all of these fixed sets. Consequently, there are symmetric orbits on each rotational torus, and if the dynamics on the torus is conjugate to an incommensurate rotation, then the symmetric orbits are dense.
Therefore using symmetric periodic orbits as limiting approximations to rotational tori seems reasonable.

\begin{table}
\centering
\begin{tabular}{c|ccc}
 $n$ &	$S_i$		&	$S_f$ & $\chi$ \\
\hline
even	&	$S_1$	&	$S_1 \circ T_{-m}$ & 	$\frac{n}{2}$ \\
	&	$S_2$	&	$S_2 \circ T_{-m}$ & 	$\frac{n}{2}$ \\
\hline
odd &	$S_1$	&	$S_2\circ T_{-m}$ &	$\frac{n+1}{2}$ \\
	&	$S_2$	&	$S_1 \circ T_{-m}$ &	$\frac{n-1}{2}$ \\
\end{tabular}
\caption{\footnotesize Initial, $S_i$, and final, $S_f$, symmetries for $(m,n)$-symmetric periodic orbits for a lift $F$ of a reversible map with reversors $S_1$ and $S_2 = F \circ S_1$ and discrete rotation symmetry $T_m$. The initial point $(x_0,z_0) \in \fix{S_i}$ maps to the point $(x_{\chi},z_{\chi}) \in \fix{S_f}$ in $\chi$ iterations.}
\label{tbl:Symmetry}
\end{table}

\subsection{Stability}\label{sec:stability}

The linear stability of a period-$n$ orbit, $(x^*,z^*) = f^n(x^*,z^*)$, is determined by its Jacobian matrix, $Df^n(x^*,z^*)$. When $f:M \to M$ is volume-preserving and $M$ is three-dimensional, the characteristic polynomial takes the form
\[
	\det(\lambda I - Df^n(x^*,z^*)) = \lambda^3 -\tau_1 \lambda^2 + \tau_2 \lambda -1 ,
\]
where the trace $\tau_1$ and second trace $\tau_2$ are defined by
\bsplit{tauSigma}
	\tau_1 &= \tr(Df^n) , \\
	\tau_2 &= \tfrac12 (\tau_1^2 -\tr(Df^{2n})) .
\esplit
There are eight generic multiplier configurations in the $\tau_1$-$\tau_2$ plane as shown in \Fig{Stability} \cite{Lomeli98}. The line $\tau_1=\tau_2$ corresponds to the existence of a multiplier $\lambda = 1$. The only ($\tau_1$,$\tau_2$) values corresponding to stable orbits lie on the segment $-1 < \tau_1=\tau_2 < 3$ where there is a conjugate pair of multipliers on the unit circle. When $\tau_1>3$, this pair becomes hyperbolic, and, when $\tau_1 < -1$, reflection hyperbolic. 

\InsertFig{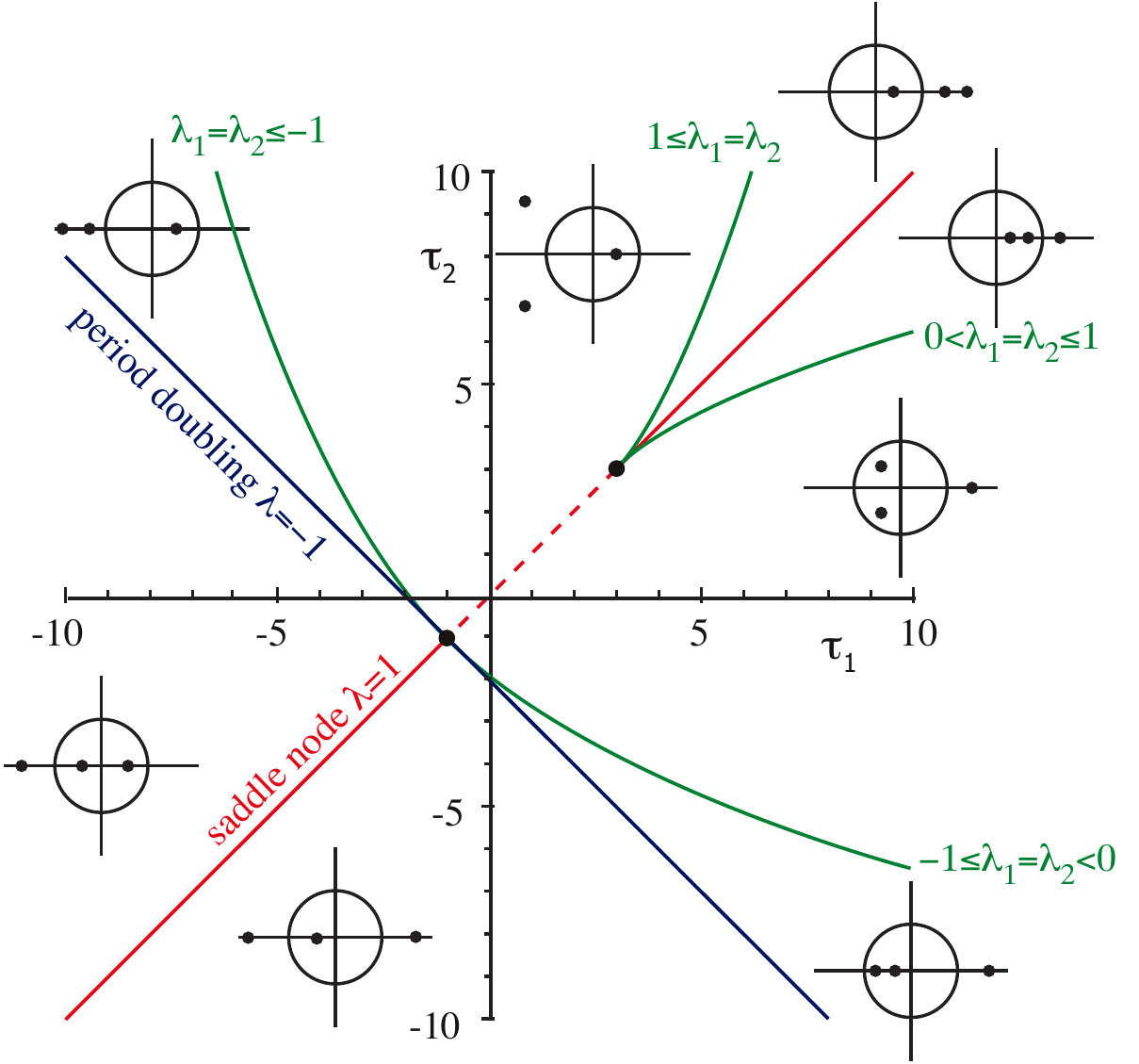}{Stability diagram for a three-dimensional, volume-preserving map depending upon the trace $\tau_1$ and second trace $\tau_2$. The eight insets show multiplier configurations in the complex-$\lambda$ plane relative to the unit circle for each of the eight stability domains.}{Stability}{3.4in}

The Jacobian $Df$ of \Eq{VPMap} is the matrix (in $2 \times 1$ block form)
\beq{Jacobian}
	Df = \begin{pmatrix} I -\eps \nabla\Omega(z') D g(x) & \nabla\Omega(z') \\
	 -\eps Dg(x) & 1
	 \end{pmatrix} .
\eeq
In this expression, $Dg$ is the row vector $(\nabla g)^T$, while $\nabla \Omega(z)$ is a column vector, so that $\nabla\Omega Dg$ is the outer product of these vectors.

Interestingly, \Eq{Jacobian} has the property that, for $n=1$,
\beq{tau}
	\tau_1 = \tau_2= 3 -\eps \nabla\Omega(z') \cdot \nabla g , \\
\eeq
for any functions $\Omega$ and $g$.
Consequently one of the multipliers of $Df$ is always $1$, and thus every fixed point has multipliers $(1,\lambda, 1/\lambda)$. For example, in \Fig{fixedPoints} the unstable fixed points are indicated by dashed ($\tau_1<-1$) and dotted ($\tau_1>3$) curves. For the parameters \Eq{StdParams}, the fixed points on the line $x_1-x_2 = \tfrac12$ are stable up to
$
	\eps = [\pi(2\beta z^*-1)]^{-1}
$
with $z^*$ given by \Eq{ZDeltaStar}. At this parameter value, $\tau_1=-1$ for the point $(\tfrac12,0,z^*)$, and as $\eps$ grows there is an interval of unstable fixed points orbits on this line. Similarly on the line $x_2 = 0$ the point $(0,0,z^*)$ first loses stability with $\tau_1 = -1$ at $\eps= \tfrac{1}{\pi}$. The fixed points on the line $x_2=\tfrac12$ have $\tau_1 > 3$ for all $\eps > 0$.

\subsection{The Residue} \label{sec:VPResidue}

We noted above that $\tau_1 = \tau_2$ for fixed points of any map of the form \Eq{VPMap}. Of course the equality $\tau_1 = \tau_2$ need not be true for $Df^n(x,z)$, so periodic orbits with periods larger than $1$ may not have a unit multiplier. However, for symmetric orbits, the property $\tau_1 = \tau_2$ follows from the conjugacy of $Df$ to $Df^{-1}$.

\begin{lem} If $\Gamma$ is a symmetric periodic orbit of a reversible, three-dimensional, volume-preserving map, then it must have a multiplier $\lambda = 1$, or equivalently, $\tau_1 = \tau_2$.
\end{lem}
\begin{proof} This follows from the more general result: each symmetric periodic orbit of a reversible diffeomorphism has reciprocal multipliers, i.e., if $\lambda$ is multiplier, then so is $\lambda^{-1}$. Indeed by \Lem{FixedSets} in the Appendix, there is a point $x^* \in \Gamma$ and a reversor $S$ such that $x^* \in \fix{S}$.  Moreover, since  $S \circ f = f^{-1} \circ S$, then $f^n = S^{-1} \circ f^{-n} \circ S$, so that the matrix $A = Df^n(x^*)$ is conjugate to $Df^{-n}(x^*)$. Since $Df^{-n}(x^*) = A^{-1}$, $A$ and $A^{-1}$ have the same spectrum.

Finally, for the three-dimensional, volume-preserving case, the product of the three multipliers is one, and since the multipliers come in reciprocal pairs, there must be a multiplier $\lambda = 1$. This is equivalent to $\tau_1 = \tau_2$.
\end{proof}

Using this result, we let $\tau = \tau_1 = \tau_2$ and define an analogue to Greene's residue \Eq{GreenesResidue} for symmetric orbits
\beq{Residue}
	R=\frac{1}{4}(3-\tau) \;, \quad \tau = \tr(Df^n(x,z)) .
\eeq
This definition was also used for a study of periodic orbits of the rotating standard map \cite{Artuso92, Tompaidis96b}---reversibility was not required for that map, since one of the three multipliers is identically equal to one for all orbits.

Just as for Greene's residue, orbits are stable when $0 < R < 1$ where they have a complex conjugate multiplier pair on the unit circle; these are analogous to elliptic orbits. The complex pair hits minus one at $R=1$ and is subsequently replaced by reciprocal pair of negative multipliers when $R>1$. When $R=0$ the orbit has a triple multiplier $\lambda = 1$, and when $R<0$ there is a reciprocal pair of positive eigenvalues; these orbits are analogous to hyperbolic orbits of an area-preserving map. 

Analytical formulae for the positions and residues of the symmetric fixed points are shown in \Tbl{per1per2}. The fixed points lie on $\fix{S_1} \cap \fix{S_2}$, recall \Eq{FixedSets}, since $m = \Omega(z^*)$. Their residues depend upon the parity of the components of $m$, which we denote by ``$e$" for even and ``$o$" for odd. There are four parities as shown in the table. 

Symmetric, period-two orbits must solve the system \Eq{PeriodTwo}.
Thus on $\fix{S_1}$, the orbit is
\[
	(0, \frac{m_1}{2}-\gamma) \mapsto (\frac{m}{2},\frac{m_1}{2}-\gamma) ,
\]
for $\delta^* = \tfrac14 \beta(m_1-2\gamma)^2-\tfrac{m_2}{2}$. Since $gcd(m_1,m_2,n) = 1$, there are three types of such orbits, depending upon the parity of $m$. Their
residues depend similarly on this parity, see \Tbl{per1per2}. A period-two orbit that begins on $\fix{S_2}$ is given by
\[
	(\tfrac12 \Omega(z^*),z^*) \mapsto (m-\tfrac12 \Omega(z^*),m_1-2\gamma-z^*) ,
\]
where the initial action must satisfy the transcendental equation
\[
	2z^* - g(\tfrac12 \Omega(z^*)) = m_1-2\gamma ,
\]
and $\delta^*$ is determined by \Eq{PeriodTwo}. The residue of these orbits does not have a simple analytical form.

\begin{table}[ht]
\centering
\begin{tabular}{@{}c|ccc@{}}
 $(m,n)$ &	$(x^*,z^*)$ & $\delta^*$	& $R$ \\	
\hline
 $(e,e,1)$ & $(0,0, m_1-\gamma)$ 	& & $\pi \eps$ \\
 $(e,o,1)$ & $(0,\frac12,m_1-\gamma)$ 	& & $0$ \\
 $(o,e,1)$ & $(\frac12,0,m_1-\gamma)$& 	
 	\raisebox{+1.5ex}[0pt]{$\beta{z^*}^2 -m_2$} & $\pi \eps(2\beta z^*-1)$ \\
 $(o,o,1)$ & $(\frac12,\frac12,m_1-\gamma)$ & & $-2\pi\beta \eps z^*$ \\ 
\hline
 $(o,o,2)$ & $(0,0,\frac{m_1}{2}-\gamma)$& & $2\pi\eps(1- 2\beta z(1- 2\pi \eps ))$ \\
 $(o,e,2)$ & $(0,0,\frac{m_1}{2}-\gamma)$& $\beta{z^*}^2-\tfrac12{m_2}$ & 
 				 $2\pi\eps(2\pi\eps +2\beta z(1- 2\pi \eps))$ \\
 $(e,o,2)$ & $(0,0,\frac{m_1}{2}-\gamma)$& &$2\pi\eps $ \\ 
\end{tabular}
\caption{\footnotesize Properties of the four types of fixed points and three types of period-two orbits beginning on $\fix{S_1}$ for $a=b=c=1$. Orbits are classified by the parity of $m$. }
\label{tbl:per1per2}
\end{table}

\section{Computing Symmetric Orbits}\label{sec:Numerics}
The computation of symmetric periodic orbits can be reduced to finding the zeros of a function $H: \bR^2 \to \bR^2$ since the orbits must begin and end on fixed sets of the reversors. Indeed, for a given fixed set, the initial angles are completely determined by $(z,\delta)$. We therefore need only determine $z_0 = z^*$ and $\delta^*$ by requiring that $x_n = x_0+m$. In addition, the numerical work can be halved: since the ``half-orbit" lands on another fixed set, recall \Tbl{Symmetry}, only $\chi \sim n/2$ iterations need to be performed.

More concretely, let $(x^i(z,\delta),z) \in \fix{S_i}$ and $(x^f(z,\delta),z) \in\fix{S_f}$ denote points on the initial and final symmetry curves of an $(m,n)$-orbit, given in \Tbl{Symmetry} using \Eq{FixedSets}. If, for example, $S_f = S_1 \circ T_{-m}$ then $x^f(z,\delta) = \tfrac12 m$, and if $S_f = S_2 \circ T_{-m}$ then $x^f(z,\delta) = \tfrac12(\Omega(z,\delta)+m)$. Denoting the length of the half-orbit by $\chi$ so that $(x_{\chi},z_{\chi}) = F^{\chi}(x^i(z,\delta),z)$, then an $(m,n)$-orbit corresponds to a zero of
\beq{NewtonFunction}
	H(z,\delta)=x_{\chi}(z,\delta)-x^f(z_{\chi}(z,\delta)) ,
\eeq
that is, to a point $(z^*,\delta^*)$ such that $H(z^*,\delta^*) = 0$.

Numerical solution of this two-dimensional system is straightforward using a Newton or pseudo-Newton method; we employ Broyden's method \cite{Dennis96}. Recall that finite difference approximations, such as those used in the secant method, are underdetermined in more than one dimension. The idea of Broyden's method is to begin with an initial approximation to the Jacobian, the identity in our case, and improve it at each iteration by taking the solution of the finite difference approximation that minimally modifies (in the Frobenius norm) the current approximation. Using double precision arithmetic, we  generally define convergence as $|H(z^*,\delta^*)|<10^{-10}$; however, orbits were often found to higher precision. A maximum of 75 Broyden iterations were performed. 

An example of the dependence of $(z^*, \delta^*)$ on $\eps$ for three low-period orbits is shown in \Fig{zDeltaVsEps}; note that the dependence on $\eps$ appears to be smooth. Orbits with periods up to $\cO(10^5)$ and moderate values of $\eps$ can be easily found. Moreover, we are able to find symmetric orbits for any coprime $(m,n)$ and any pair $S_i$, $S_f$ as listed in \Tbl{Symmetry}.

It is important to note that errors can be compounded in these calculations. While the half-orbit error was always smaller than $10^{-10}$, the error for the full orbit,
\beq{OrbitError}
	\kappa \equiv | x_n - x_0 -m| + |z_n-z_0| ,
\eeq
was often larger, but generally no more than $10^{-8}$. Of course when the orbit was stable or nearly stable, $R = \cO(1)$, the error $\kappa$ was typically smaller. 

Unstable orbits are difficult to find because the derivative of $H$ becomes large and the algorithm becomes extremely sensitive to the initial guess. This can be partially obviated by using continuation from the trivial case $\eps = 0$ where the initial conditions are given by \Eq{ZDeltaStar}. We found quadratic extrapolation to be sufficient to predict subsequent initial conditions. Extrapolation typically requires 10-15 fewer iterates of the Broyden method, allows larger steps in $\epsilon$, and converges for larger residue values. 

To maintain convergence, the extrapolation step size must decrease as the orbit period grows. To start, the simple choice $\Delta\eps= 0.1 n^{-1}$ yields few convergence failures. If successful, the step size is increased by half. If unsuccessful, the step size is decreased by a third. Though this process generally results in more failed steps than successful ones, it proved far faster than a method with a fixed step size.

\InsertFigTwo{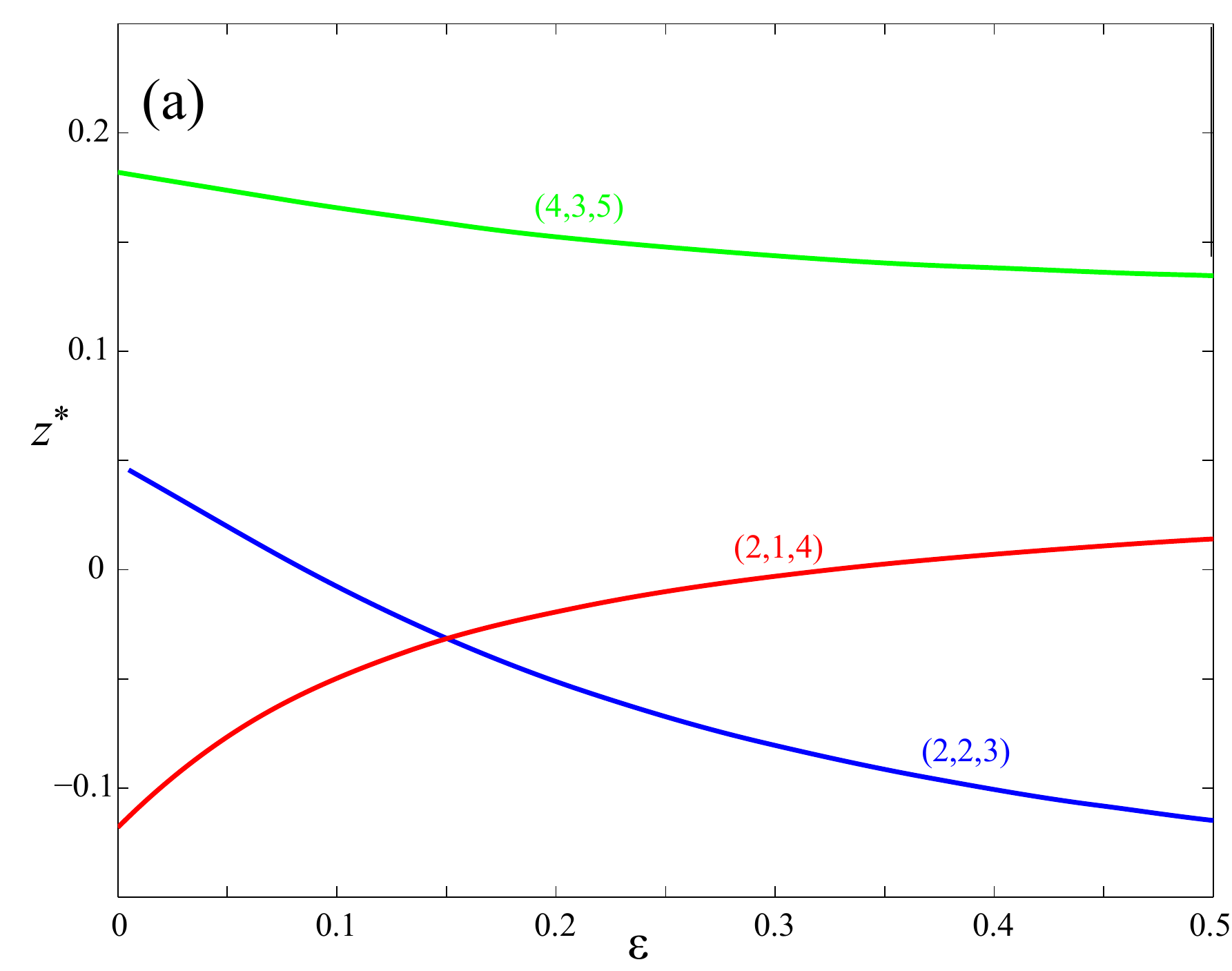}{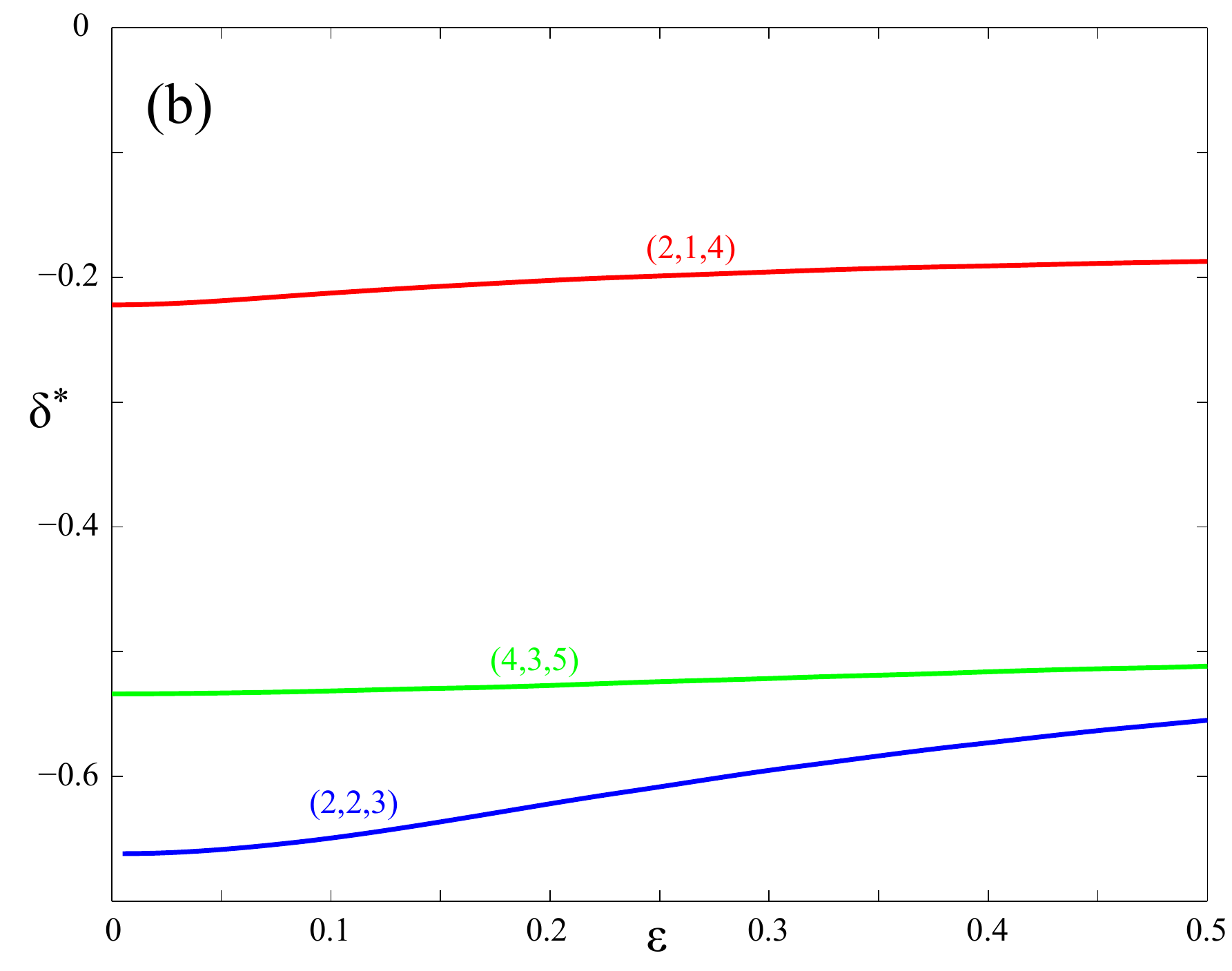}{Numerically computed $(z^*,\delta^*)$ for three low-period orbits beginning on $\fix{S_1}$ as a function of $\eps$ using a step size of 0.005}{zDeltaVsEps}{3.5in}

\section{Computing Residues}\label{sec:NumericalResidues}
Given numerical values of $(z^*, \delta^*)$ for an $(m,n)$-orbit, the residue \Eq{Residue} is obtained by multiplying the successive Jacobian matrices along the orbit to compute $\tr(Df^n)$. Every $(m,n)$-orbit has zero residue at $\eps =0$ and we observe that $R$ grows or decreases smoothly with $\eps$. The behavior is initially monotone, but for intermediate values it often oscillates, as can be seen for several low-period orbits with $(x^i,z^i) \in\fix{S_1}$ in \Fig{FourResidues}. For example, the residue of the $(2,1,4)$ orbit (red curve) initially becomes negative, reaching a minimum near $\eps = 0.04505$ and subsequently grows monotonically, passing $R=1$ near $\eps = 0.1146$. 
The residue of the $(1,3,4)$ orbit (black curve), on the other hand, is initially positive, but reaches a maximum near $\eps=0.09815$, and then decreases monotonically thereafter. The residues of higher period orbits can oscillate several times, but always appear to grow monotonically for large $\eps$. 

\InsertFig{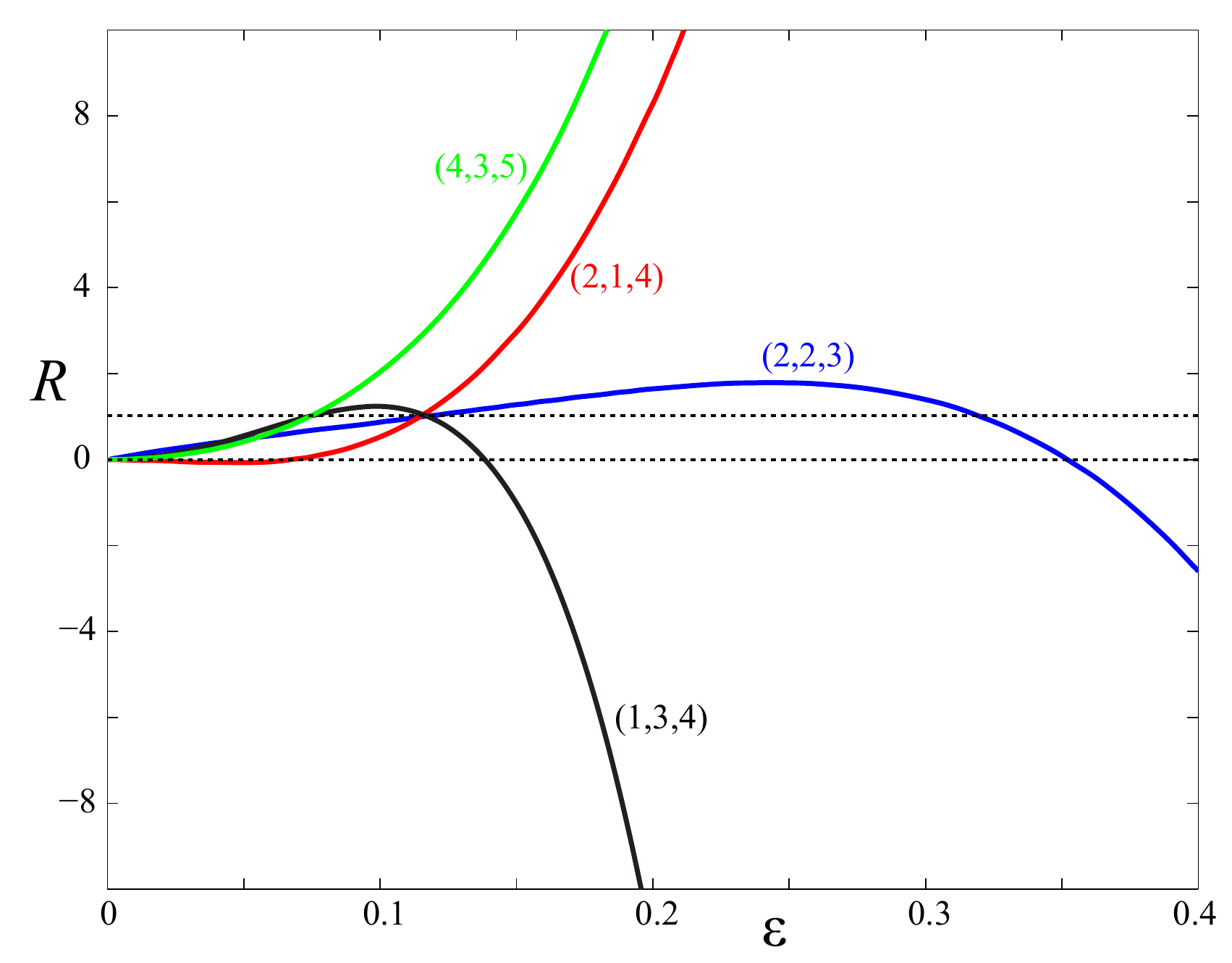}{Residue as a function of $\eps$ for four $(m,n)$-orbits of \Eq{VPMap} that start on $\fix{S_1}$. The bifurcation values $R=0$ and $R=1$ are indicated by dashed lines.}{FourResidues}{4.5in}

Note that the residue for fixed points and period-two orbits either grows linearly or quadratically for small $\eps$ (except for the $(e,o,1)$ orbits which have zero residue), recall \Tbl{per1per2}. Surprisingly, we observe that linear or quadratic growth holds for higher periods as well, some samples are shown in \Fig{AsymptoticResidue}. There does not appear to be a simple relation between the period or type of orbit and the small-$\eps$ asymptotic behavior. For example, we observed the growth rates:
\begin{align*} 
R = \left\{ \begin{array}{lcl}
	 \cO(\eps) &:& (1,1,3), (2,2,3), (2,3,3), (3,2,3), (3,3,4), (1,4,4), (3,4,4), 
		(4,3,4), (1,1,5), \ldots \\
	 \cO(\eps^2) &:& (1,2,3), (2,1,3),(2,1,4), (1,3,4), (3,2,4), (1,3,5), (3,2,5),
		(4,2,5), (4,3,5), \ldots 
\end{array} \right. .
\end{align*}
when $\eps \ll 1$, independent of the initial symmetry lines.

This asymptotic behavior contrasts with the behavior of the standard map, where $R = \cO(\eps^{n})$ for $\eps \ll 1$ \cite{Greene79}. An explanation for our observations could follow from a generalization of the Poincar\'e-Birkhoff theorem to the volume-preserving case, but we know of no such result.\footnote
{For example, Cheng and Sun's generalization of Poincar\'e-Birkhoff applies to invariant circles, not periodic orbits \cite{Cheng90b}.}

\InsertFig{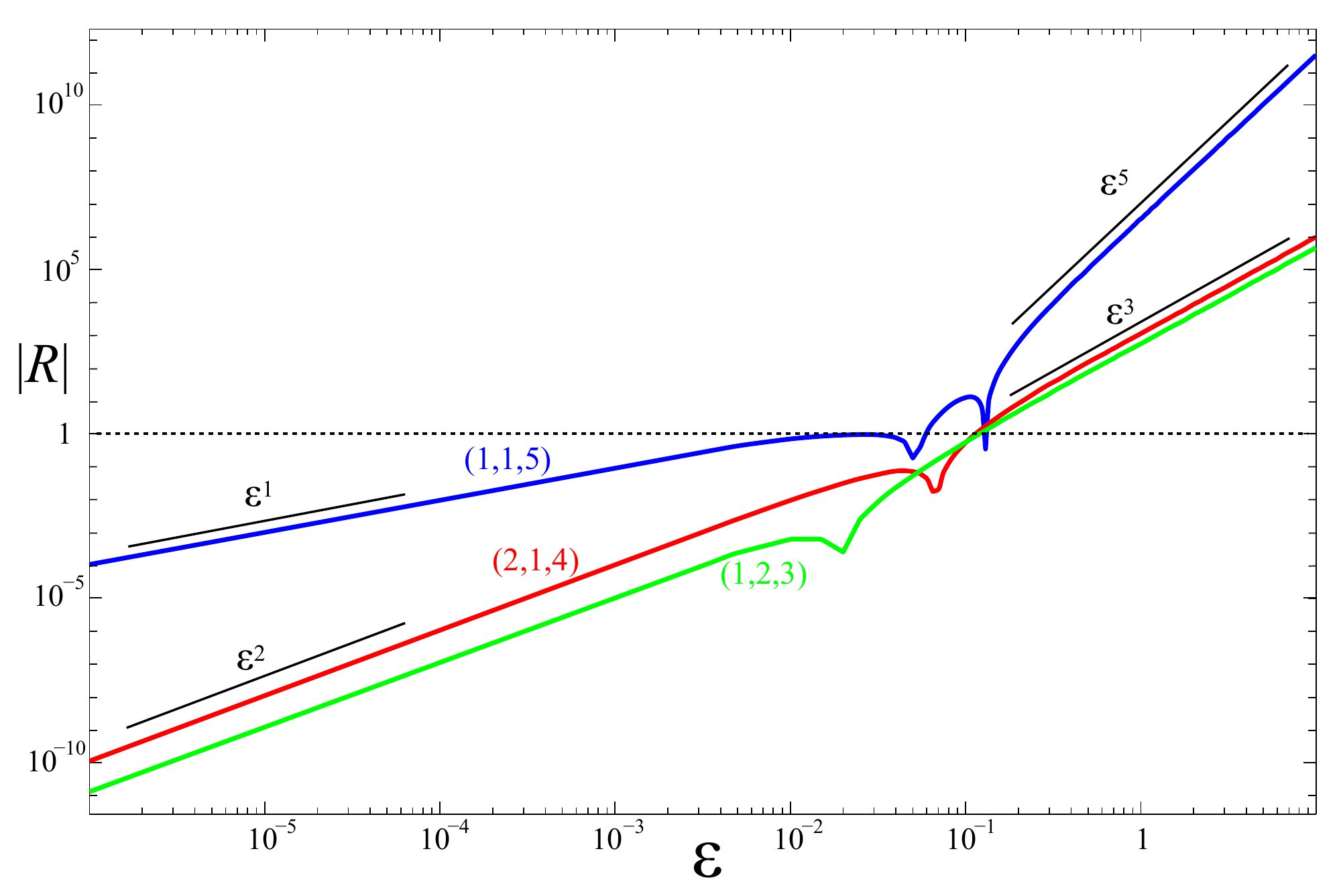}{Residue as a function of $\eps$ for three orbits of \Eq{VPMap} that start on $\fix{S_1}$ plotted on a log-log scale. The cusps correspond to points where $R$ changes sign. The asymptotic behavior is indicated by the line segments.}{AsymptoticResidue}{5.5in}

Greene also observed that for the standard map, $R = \cO(\eps^{n})$ for $\eps \gg 1$. We similarly observe that most of the residues of \Eq{VPMap} grow as $\eps^n$, three examples are shown in \Fig{AsymptoticResidue}. However, there are exceptions: every symmetry line has a class of orbits---those listed in \Tbl{AsyLarge}---for which the residue grows as $\eps^{n-1}$. Moreover, on $\fix{S_1\circ T_{0,1}}$ the residues of all orbits have this lower growth rate. 

\begin{table}[bht]
\begin{center}
\begin{tabular}{@{} c| c| c| c| c| c| c| c @{} }
 $S_1$  & $S_1\circ T_{0,1}$ & $S_1\circ T_{1,0}$ & $S_1\circ T_{1,1}$ & $S_2$  & $S_2\circ T_{0,1}$ & $S_2\circ T_{1,0}$ & $S_2 \circ T_{1,1}$ \\ \hline
 $(e,o,e)$ &  All  & $(o,o,e)$  & $(o,e,e)$  & $(e,o,o)$ &  $(e,e,o)$ & $(o,o,o)$ & $(o,e,o)$ 
 \end{tabular}
 \caption{\footnotesize Classes of orbits for which $R = \cO(\eps^{n-1})$ for $\eps \gg 1$, grouped by symmetry line. Orbits not listed have $R = \cO(\eps^n)$ for $\eps \gg 1$.}
\label{tbl:AsyLarge}
\end{center}
\end{table}

As we observe in \Fig{FourResidues}, the sign of the residue varies with 
the orbit and with $\eps$.
Greene and MacKay found that the standard map has a special symmetry fixed set, the so-called \emph{dominant} set, on which every orbit of the standard map has positive residue \cite{MacKay92a}. This holds as well in other two-dimensional examples \cite{Roberts92}.
To our knowledge no such result is known for higher-dimensional maps, though there is some numerical evidence of a dominant symmetry for the 4D Froeshl\'e map \cite{Kook89}. 

To determine whether the map \Eq{VPMap} has a dominant symmetry, we looked at a number of $(m,n)$-orbits of each symmetry and parity type. The results are summarized in \Tbl{lowperiod}, which indicates the stability of the orbit on each of the eight symmetry fixed sets. Unfortunately, we were unable to find any systematic organization of residue signs on any fixed set. One pattern that seems to hold is that for small, positive $\eps$, there appear to be two elliptic, $0 < R < 1$, and two hyperbolic, $R < 0$, orbits for each $(m,n)$; these are labeled $e_i$ and $h_i$ for $i=1,2$ in \Tbl{lowperiod}. This does not hold, however, for large $\eps$: the asymptotic sign of the residue as $\eps \to \infty$ is indicated by the $\pm$ signs in the table.

\begin{table}[ht]
\begin{center}
\begin{tabular}{@{} c | c c c c c c c c @{} }
Orbit & $S_1$ & $S_1\circ T_{0,1}$ & $S_1\circ T_{1,0}$ & $S_1\circ T_{1,1}$ & $S_2$ & $S_2\circ T_{0,1}$ & $S_2\circ T_{1,0}$ & $S_2 \circ T_{1,1}$ \\ \hline
(2,2,3) & $e_1-$ & $h_1-$ & $h_2-$ & $e_2+$ & $e_1-$ & $h_1-$ & $h_2-$ & $e_2+$ \\ 
(4,2,5) & $h_1-$ & $h_2+$ & $e_2+$ & $e_1-$ & $h_1-$ & $h_2+$ & $e_2+$ & $e_1-$ \\ 
\hline
(2,1,4) & $h_1+$ & $h_1+$ &$e_1+$ & $e_1+$ & $h_2-$ & $h_2-$ & $e_2-$ & $e_2-$ \\ 
(4,3,4) & $e_1+$ & $e_1+$ & $h_2-$ & $h_2-$ & $h_1+$ & $h_1+$ & $e_2-$ & $e_2-$ \\ 
\hline
(2,1,3) &$e_1+$ & $h_1+$ & $e_2-$ & $h_2-$ & $h_1+$ & $e_1+$& $h_2-$ & $e_2-$ \\
(4,3,5) &$e_1-$& $h_1+$ & $e_2-$& $h_2+$ & $h_1+$ & $e_1-$& $h_2+$& $e_2-$\\ 
\hline
(1,4,4)& $h_1+$ & $e_1+$ &$h_1+$ & $e_1+$ &$e_2-$ & $h_2-$ & $e_2-$ & $h_2-$ \\
(3,2,4)& $e_1+$ & $h_1+$ &$e_1+$ & $h_1+$ &$h_2+$ & $e_2+$ & $h_2+$ & $e_2+$ \\ \hline
(1,2,3)& $h_1+$ & $h_2-$ & $e_1+$ & $e_2+$ & $e_1+$ & $e_2+$ & $h_1+$ & $h_2-$ \\
(3,2,5)& $e_1+$ & $h_1+$ & $h_2+$ & $e_2+$ & $h_2+$ & $e_2+$ & $e_1+$ &$h_1+$ \\ \hline
(1,3,4)&$e_1-$ & $e_2-$ &$e_2-$ &$e_1-$ &$h_1-$ &$h_2-$ &$h_2-$&$h_1-$ \\
(3,3,4)&$e_1-$ &$h_1-$ &$h_1-$ &$e_1-$&$e_2+$&$h_2-$&$h_2-$&$e_2+$ \\ 
\hline
(1,1,3)&$e_1+$ &$h_1-$ & $h_2-$&$e_2-$ &$e_2-$ & $h_2-$&$h_1-$ &$e_1+$ \\
(1,3,5)&$h_1+$ &$e_2-$ &$e_1-$ &$e_2-$ &$e_2-$ & $e_1-$&$e_2-$ &$h_1+$ \\ 
\end{tabular}
\caption{\footnotesize Signs of the residues of symmetric orbits of each parity on each of the eight symmetry fixed sets for \Eq{VPMap} with parameters \Eq{StdParams}. Orbits are labeled by stability for small positive $\eps$ (``$e$" for elliptic and ``$h$" for hyperbolic), and by the residue sign ($\pm$) for large positive $\eps$. For small $\eps$ there are two hyperbolic and two elliptic orbits for each $(m,n)$, labelled by subscripts $1$ and $2$.
}
\label{tbl:lowperiod}
\end{center}
\end{table}

The accuracy of the residue computations is less than that of the orbit itself because multiplication of the Jacobian matrices along the orbit gives rise to additional error. Roughly speaking, for fixed matrices $A$ and $B$ and $\kappa \ll 1$,
\[
	\tr((A+\kappa B)^n) \approx \tr(A^n) + \mathcal{O}(n\kappa\mu^{n-1})
\]
where $\mu$ is the largest eigenvalue of $A$. For the residue computation we can think of $\mu$ as an estimate of the Lyapunov multiplier or mean residue of the orbit, and thus the relative error in the residue is of order $\frac{n\kappa}{\mu}$.
If $\kappa$ were fixed and $\mu = \cO(1)$, the error would grow linearly with the period. However, as the period grows, the orbit error $\kappa$, \Eq{OrbitError}, increases because of the floating point errors in the computation of \Eq{NewtonFunction}. 

This error estimate for the residue computations is shown in \Fig{ResidueError}(a) for a period $5842$ orbit. When $\eps < 0.023$, $|R| <\cO(10^{-7})$, and the error makes the residue essentially uncomputable; however, the main point is that all of the multipliers for small $\eps$ are essentially $1.0$. When the orbit loses stability, near $\eps = 0.025$, it does so rapidly reaching $R=10,010$ at $\eps= 0.0265$, and the estimate indicates that when $10^{-7} < |R| < 10^{4}$, the relative error in $R$ is small. Note that even when $R \sim 10^4$, the Lyapunov multiplier of this orbit is $\mu \sim R^{1/n} = \cO(1)$; consequently the error in the residue is still of order $n \kappa$.

The residue as a function of period for $35$ orbits for $\eps = 0.01$ is shown in \Fig{ResidueError}(b), together with the error estimate. This particular sequence of orbits corresponds to the Farey sequence of approximants of the ``spiral mean" rotation vector that will be studied in \Sec{SpiralMean}. For this $\eps$, these orbits become increasingly stable as the period grows. While we are not able to obtain an accurate value for the residue when $n > 300$, we can nevertheless be sure that it is very small: the multipliers of these orbits sit very close to the point $\tau=3.0$. We will argue in \Sec{CriticalTorus} that this sequence of periodic orbits appears to converge to an invariant torus.

\InsertFigTwo{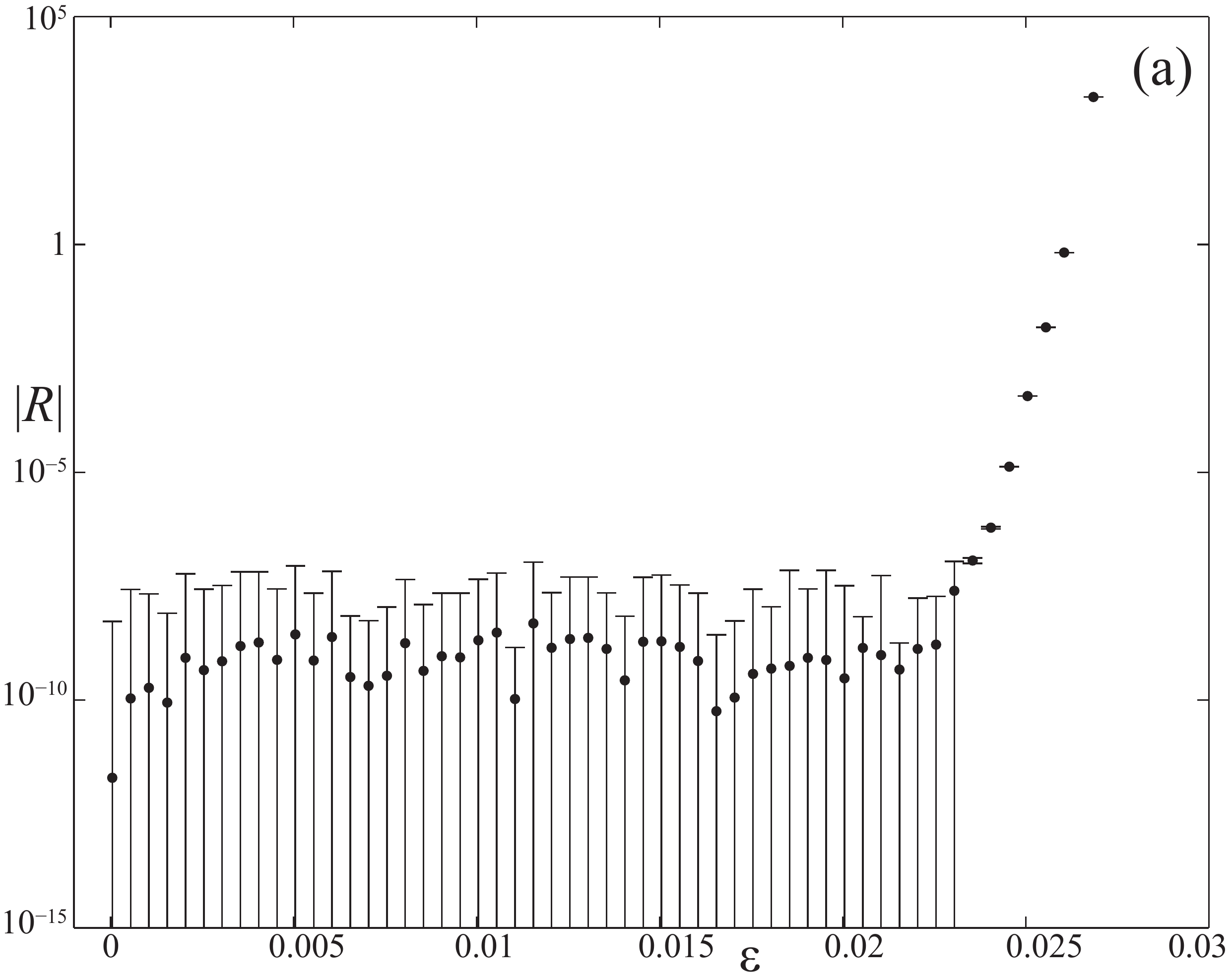}{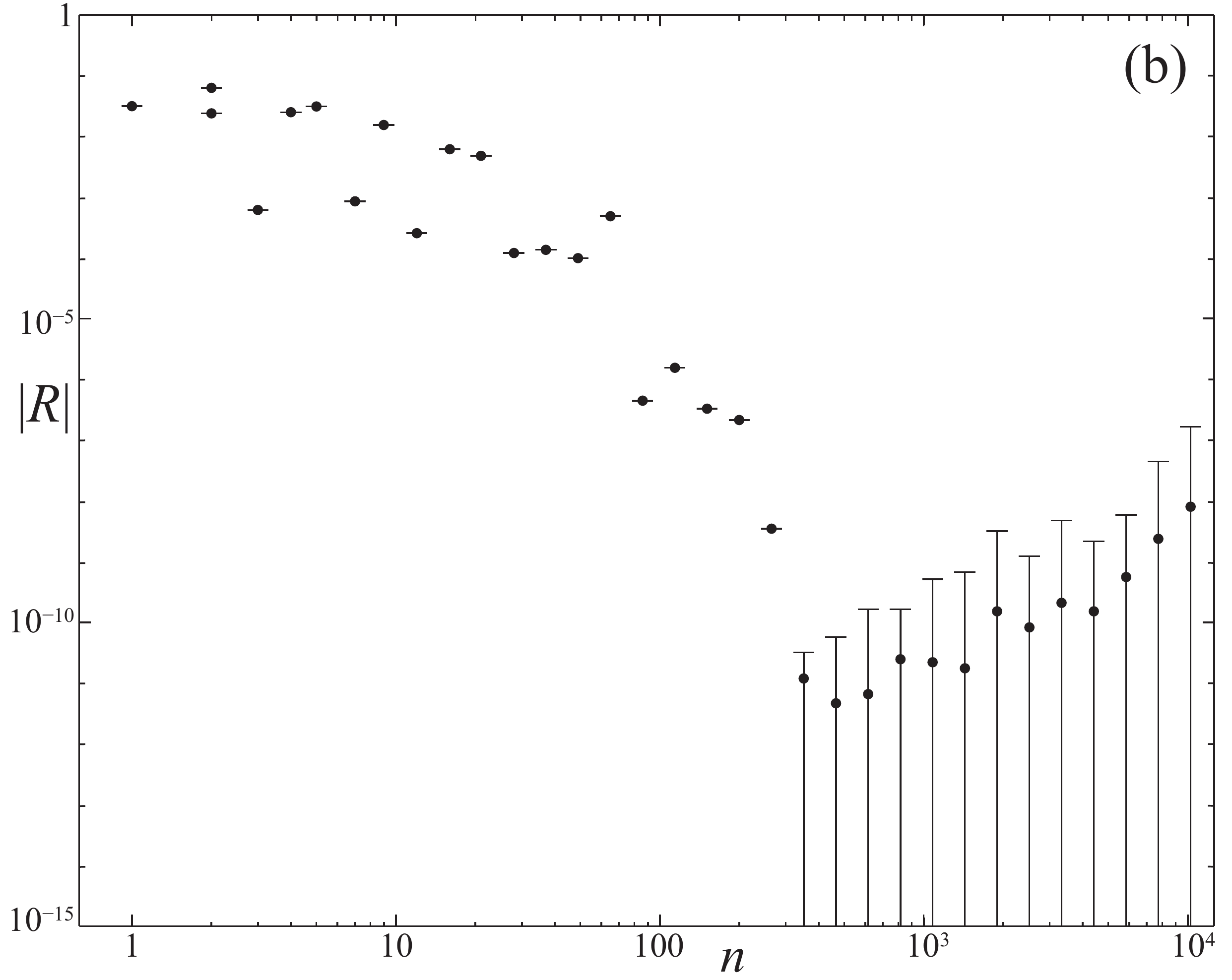}{(a) Residue of the
$(1897, 4410, 5842)$, $\fix{S_1}$ orbit as a function of $\eps$. (b) Residues of a sequence of $\fix{S_1}$ orbits with periods $1$ to $17991$ (the orbits of \Tbl{SpiralMeanEpsCR}) with $\eps = 0.01$. The bars indicate the estimated error $\pm n\kappa$.}{ResidueError}{3.4in}

\section{Farey Sequences}\label{sec:FareySequences}

As we discussed in \Sec{ResidueCriterion}, the application of Greene's residue criterion requires finding a suitable sequence of $(m,n)$-orbits whose rotation numbers converge to a chosen Diophantine rotation vector. For the $d=1$ case, these can be systematically obtained either using continued fractions---for which the successive convergents \Eq{Convergents} are ``best approximants"---or the Farey tree---which gives, in addition to the convergents, intermediate approximants.

There is no multi-dimensional generalization of the continued fraction algorithm that is guaranteed to give exactly the sequence of best rational approximants to a given irrational vector, though there are strongly convergent algorithms \cite{Lagarias94, Khanin07}. In lieu of these, we will use Kim and Ostlund's generalization of the Farey tree to generate rational approximations to incommensurate rotation vectors \cite{Kim86}.

In addition, it is not at all clear what Diophantine vector may be an appropriate choice to replace the golden mean, \Eq{GoldenMean}, which appears to give the most robust invariant circle in the standard map. Since, as we recall below, integral bases of degree-$(d+1)$ algebraic fields can be chosen to make Diophantine vectors, it seems natural to conjecture that some cubic field will lead to robust two-tori. We will investigate vectors from the so-called ``spiral mean" field. 

\subsection{Generalized Farey Tree}\label{sec:KimOstlund}

The Farey tree is a recursive algorithm for generating all rationals between a given initial pair \cite{Hardy79}. To do this, begin by regarding each $\omega \in \bR_+$ as the direction $(\omega,1)^T \in \bR_+^2$ and each positive rational $\tfrac{m}{n}$ as the direction $(m,n)^T$. Conversely, for each vector $(u,v)^T \in \bR_+^2$, there is an associated rotation number $\omega = u/v$. The problem of approximation is to then find a sequence of vectors $(m_\ell,n_\ell)^T$ that approach $(\omega,1)^T$ in the sense that the two directions become parallel, i.e., $|n_\ell\omega-m_\ell| \to 0$ as $\ell \to \infty$. 

The root of the Farey tree consists of the two vectors, $(0,1)^T$ and $(1,0)^T$, said to be at \emph{level} zero. The interior of the cone of directions spanned by the positive linear combinations of these vectors thus represents all positive numbers. The set of directions can be mapped onto the segment of the line $u+v=1$ in the positive quadrant, and each direction in the ``level-zero" cone corresponds to a point on this segment as sketched in \Fig{FareyTree}. The next level is obtained with the mediant operation $\oplus$,
\[
	\begin{pmatrix} m \\n \end{pmatrix} \oplus \begin{pmatrix} p \\ q \end{pmatrix}
	 = \begin{pmatrix} m+p \\n+q \end{pmatrix},
\]
which is simply vector addition applied to the bounding vectors.
Thus at level one there is one new \emph{daughter} vector, $(1,1)^T = (0,1)^T \oplus (1,0)^T$. The daughter divides the level-zero cone into two ``level-one" cones
\[
	C_l = \begin{pmatrix} 1 & 0 \\ 1 & 1 \end{pmatrix} , \quad
	C_r = \begin{pmatrix} 1 & 1 \\ 0 & 1 \end{pmatrix} ,
\]
where the columns represent the bounding vectors. 
This process can be repeated: each level-$\ell$ cone is divided by forming the mediant of its boundary pair. Thus the level-one cone $C_l$ becomes the two cones $C_lC_l$ and $C_lC_r$, and there are $2^\ell$ cones at level $\ell$. 

The Farey tree is then the binary graph obtained by connecting each rational vector at level $\ell \ge 1$ with its two daughters at the next level. Each vertex has a unique \emph{path}, $p \in \{l,r\}^{\bN}$: the path to $(1,1)^T$ at level one is empty, and $p=l$ for the daughter $(1,2)^T$ to the ``left" and $p=r$ for the daughter $(2,1)^T$ to the ``right". Each level-$\ell$ parent has two level-$(\ell+1)$ daughters, whose paths are obtained by appending an $l$ or $r$, respectively, to the parent's path. The vector corresponding to a path, $p = i_1\,i_2\,\ldots\, i_\ell$, $i_k \in \{l,r\}$ is then
\[
	\begin{pmatrix} m \\ n \end{pmatrix} = C_{i_1} C_{i_2} \ldots C_{i_\ell} 
		\begin{pmatrix} 1 \\ 1 \end{pmatrix} .
\]
\InsertFig{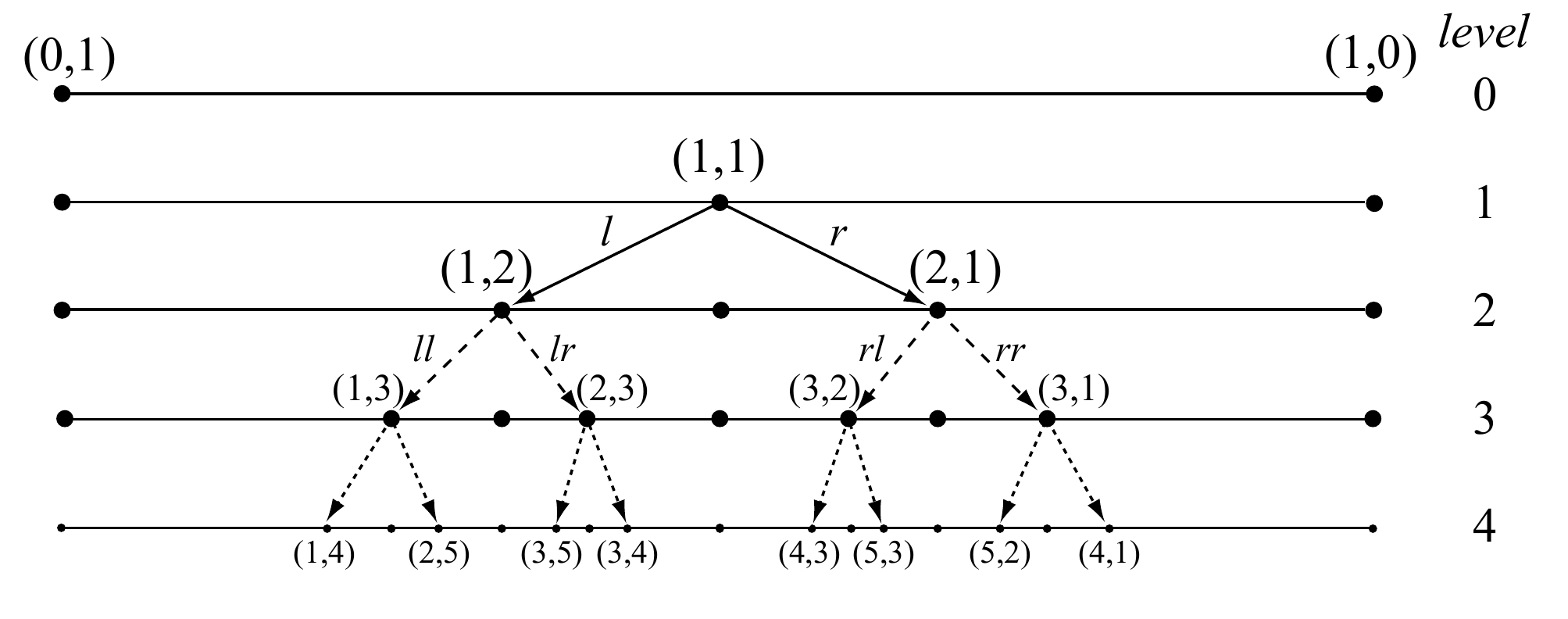}{First five levels of the Farey tree construction as the set of directions in the first quadrant projected onto the segment $x+y=1$.}{FareyTree}{4.5in}

Since the matrices $C_l$ and $C_r$ are unimodular, the parallelogram formed from the two vectors bounding each cone has unit area. A consequence is that every vector on the Farey tree is coprime, $\gcd(m,n) = 1$. Moreover, each rational has a finite Farey path, and every irrational $\omega$ is obtained as the limit of an infinite path on the tree. For example, the golden mean has path $\phi = rlrlrlrl\ldots \equiv \overline{rl}$. There are also infinite paths that limit on each rational from above and below, these are the paths with infinite tails of $l$'s and $r$'s, respectively. There is also a simple relation between the continued fraction expansion and the Farey path, see e.g., \cite{Hardy79}.

Kim and Ostlund generalize the Farey tree to higher dimensions by regarding each vector $\omega \in \bR_+^2$ as a direction $(\omega, 1)^T \in \bR_+^3$ \cite{Kim86}. We will take the level-zero cone to be the positive octant, generated by positive linear combinations of the triplet $(1,0,0)^T$, $(0,1,0)^T$, $(0,0,1)^T$. The intersection of the cone $(u,v,w)^T \in \bR_+^3$ with the plane $u+v+w=1$ is the equilateral triangle sketched in \Fig{GFT}(a). Somewhat arbitrarily, the level-one child is defined to be the mediant of two of the bounding, parent vectors; we choose $(1,1,0)^T = (1,0,0)^T \oplus (0,1,0)^T$. This creates a pair of level-one cones bounded by the child and the two parents that form an independent triplet, that is the cones 
\beq{GFTMatrices}
	C_l=\begin{pmatrix}
			0& 0 & 1 \\
			1 & 0 & 1 \\
			0 & 1 & 0 
		\end{pmatrix} , \quad
	C_r=\begin{pmatrix}
			0 & 1 & 1 \\
			0 & 0 & 1 \\
			1 & 0 & 0 
		\end{pmatrix} .
\eeq
As shown in \Fig{GFT}(a), the line connecting the unused parent, $(0,0,1)^T$ to the child, $(1,1,0)^T$ divides the equilateral triangle into ``left" and ``right" triangles. To continue the process, a new child in each cone is obtained by applying the mediant operation to the two oldest parents (those generated from the earliest levels), and each cone is divided into two, using the resonance line connecting the child and its youngest parent. The construction up to level three is shown in \Fig{GFT}(a). 

Projecting the cones onto the plane $w=1$ gives the frequency plane, as shown in \Fig{GFT}{b}. Each point $\omega$ in this plane represents a direction $(\omega,1)$; here we show only rotation vectors in the unit square $[0,1]\times[0,1]$; these correspond to directions in the cones $rrl$ and $llr$ of the full tree.

\InsertFigTwo{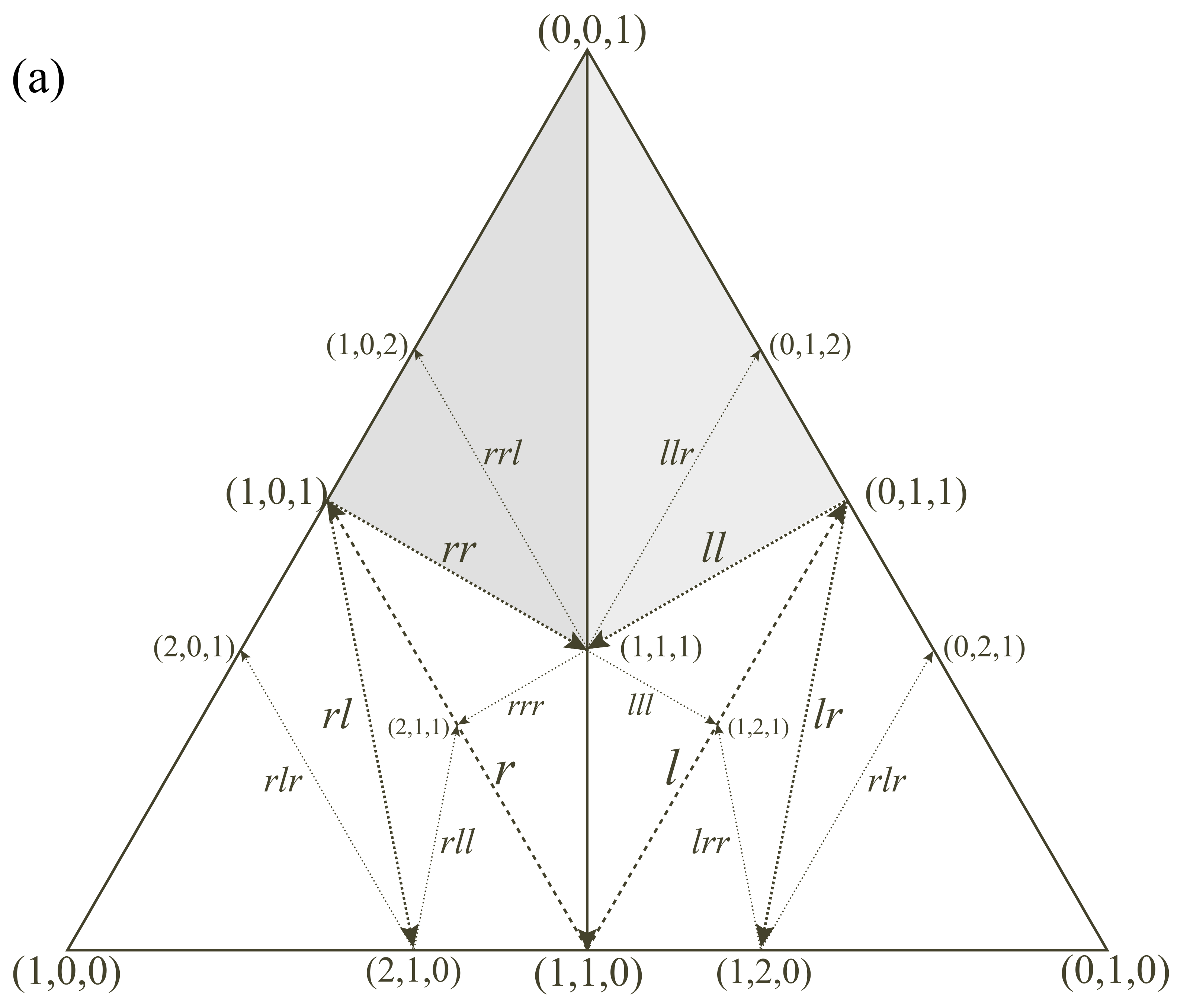}{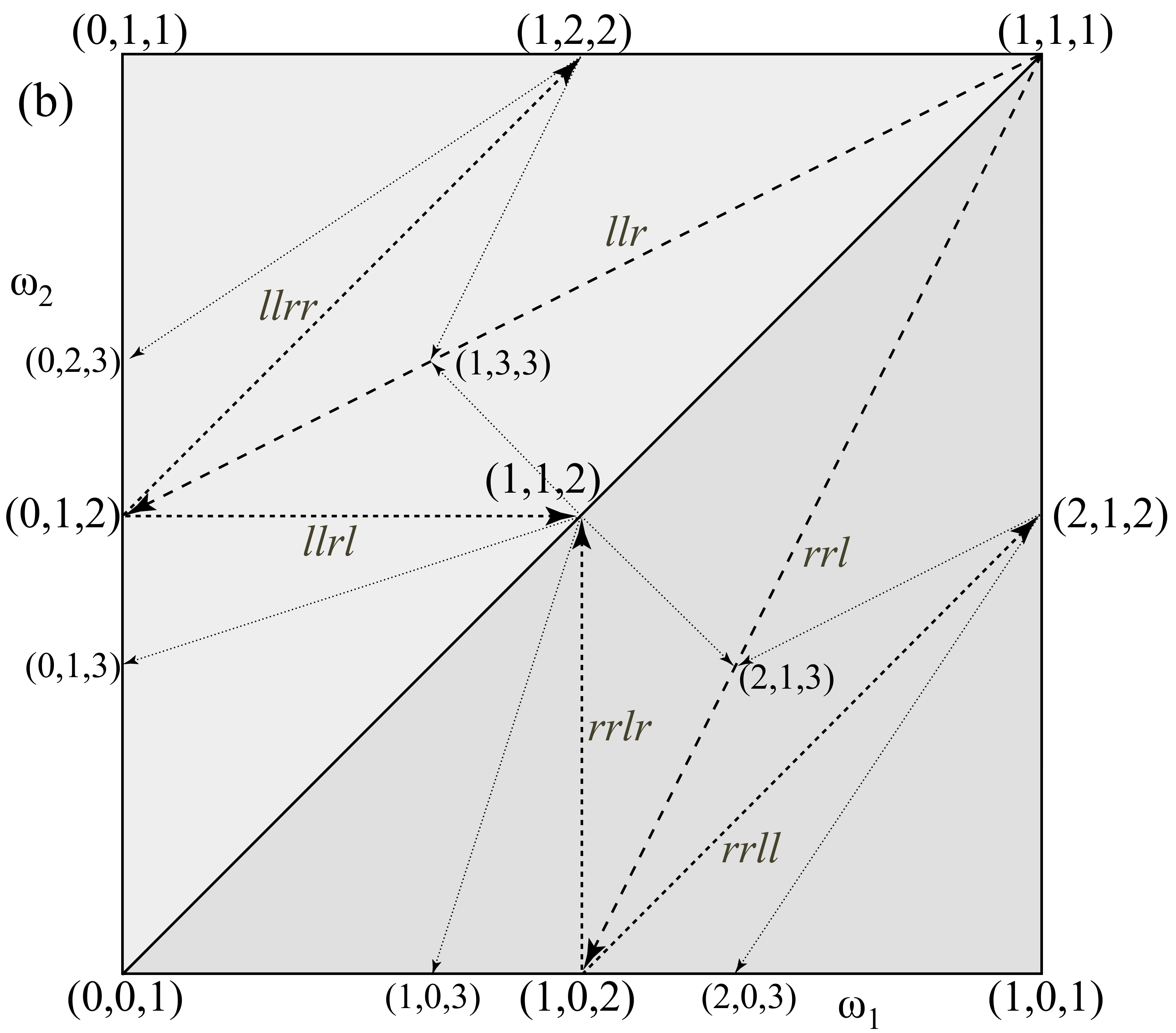}{(a) First three levels of the generalized Farey tree projected on the plane $x+y+z=1$ in homogeneous coordinates.
The two shaded triangles correspond to rotation vectors in $[0,1]^2$. (b) Projection of the two level-three cones, $rrl$ and $llr$, onto the plane $w=1$, and three additional levels of the tree.}{GFT}{3.4in}

This binary, recursive procedure assigns a path $p \in \{l,r\}^\bN$ to each vertex below level one by following the sequence of dividing lines from $(1,1,0)^T$: a turn to the ``right" appends an $r$ to the path and a turn to the left, an $l$. Thus, for example the path $rrr$ corresponds to $(2,1,1)^T$ and $lrl$ to $(0,2,1)^T$. Just as for the Farey tree, the vector corresponding to a path $p$ is
\beq{GFTPath}
	\begin{pmatrix} m_1 \\ m_2 \\ n \end{pmatrix} 
	 = C_{i_1}C_{i_2} \ldots C_{i_\ell}
		\begin{pmatrix} 1 \\ 1 \\ 0\end{pmatrix} , 
\eeq
using the matrices \Eq{GFTMatrices}.

Since the matrices $C_i$ are unimodular, the three vectors that bound each cone define a parallelepiped with volume one. A consequence is that every vector on the tree is coprime, $\gcd(m_1,m_2,n) = 1$. Moreover it can be shown that every coprime vector in the positive octant has a finite path on the tree. One defect of this construction, however, is that every integer vector in the interior of the positive octant has two finite paths; for example, $rrlll$ and $rrlrr$ both lead to $(2,1,3)^T$.\footnote
{If $(m,n)$ lies on a level-$\ell$ dividing line then its two paths are those obtained by exchanging $l$ and $r$ for all entries beyond level $\ell$.}
Just as for the Farey tree, however, every incommensurate direction has a unique, infinite path. There are also infinite paths that limit on rational directions; for example, any path $p = h\overline{lr}$ with a periodic $lr$ tail and arbitrary head $h$.

\InsertFig{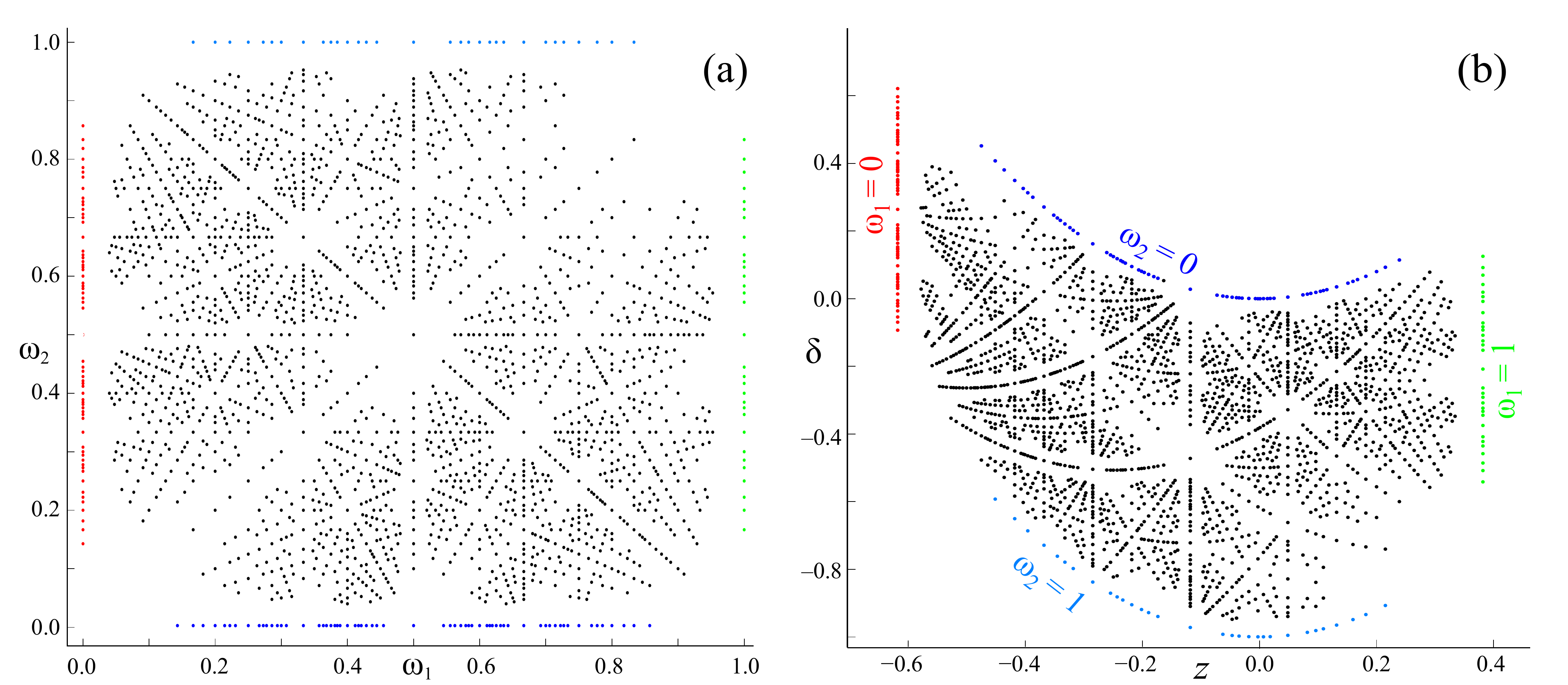}{(a) The 2140 unique rotation vectors that give $(\omega_1,\omega_2) \in [0,1]^2$ on the Kim-Ostlund tree up to level $13$. (b) The corresponding values $(z^*(\omega),\delta^*(\omega))$ for the $(m,n)$-orbits of \Eq{VPMap} for $\eps=0$. Rotation vectors on the four lines $\omega_{1,2} = 0, 1$ are colored and labeled in (b).}{level13GFT}{6.5in}

Accounting for the duplication, there are
\[
	 2^\ell + [5- (-1)^l] 2^{\floor{\frac{\ell}{2}}-1}+1
\]
unique directions on the tree up to level $\ell$ (including the initial triplet). In \Fig{level13GFT}(a) we show the $2140$ unique rotation vectors in the $rrl$ and $llr$ cones up to level $13$. The boundaries of each cone at each level correspond to resonances
$p \cdot \omega = q$, and these give rise to the isolated lines in the figure. Just as for the Farey tree, the sampling of rotation vectors is nonuniform, tending to avoid low-order resonances.

Each of the rotation vectors in \Fig{level13GFT}(a) corresponds to a periodic orbit of \Eq{VPMap} at $\eps = 0$. The $(z^*, \delta^*)$ values, given by \Eq{ZDeltaStar}, for these orbits are shown in \Fig{level13GFT}(b). Note that the transformation $\omega \mapsto (z^*,\delta^*)$ turns each line $\omega_1 = const$ into a parabola, and that since the map $\Omega(z,\delta)$, \Eq{Omega}, is orientation reversing, the upper and lower boundaries of the square are flipped.

Each of the points in \Fig{level13GFT}(b) continues to an $(m,n)$-orbit of the map \Eq{VPMap} on each symmetry line for $\eps \neq 0$. The orbits on $\fix{S_1}$ for $\eps = 0.02$ are shown in \Fig{omegaEps02}. The points are colored to indicate their residues: orbits with the smallest residues are light (yellow), those with $|R| \approx 1$ are dark (red), and those with $|R|>1$ are black (small points). The corresponding ($z^*,\delta^*)$ values for each of these orbits, \Fig{omegaEps02}(b), are shifted from their $\eps = 0$ values so as to widen the gaps around the low-order resonance curves. In phase space this correlates with the growth of tubes of secondary tori for each resonance, and these are largest for the forced resonances, $p = (0,1)$, $(1,0)$, and $(1,-1)$, of the force \Eq{Force} \cite{Dullin12}. Similar behavior has been observed in action space for symmetric orbits of the four-dimensional Froeshl\'e map \cite{Kook89}.

The lightest regions in \Fig{omegaEps02} correspond to groupings of the most stable orbits. As we will see in \Sec{ResidueSpiralMean}, these regions also contain the most robust tori.

\InsertFig{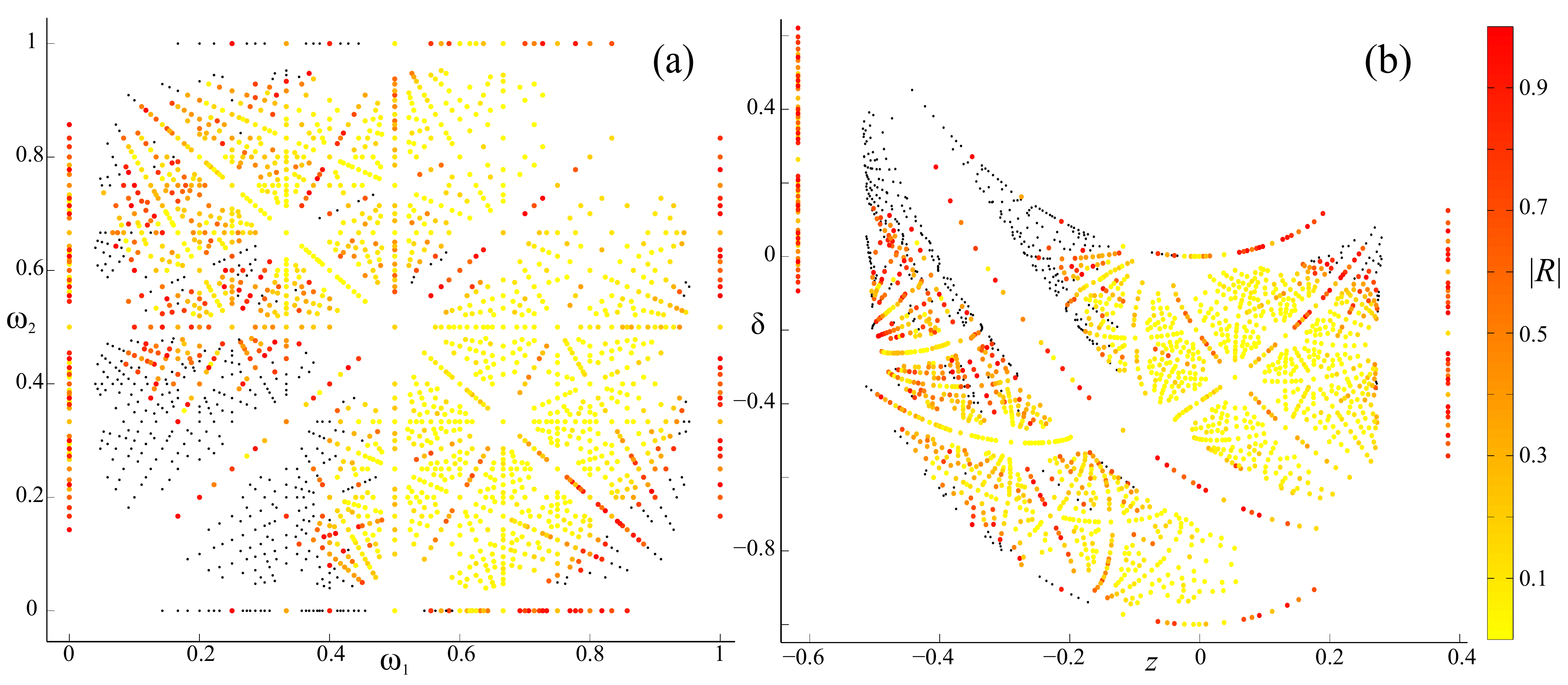}{Periodic orbits of \Eq{VPMap} on $\fix{S_1}$ for $\eps= 0.02$ with rotation vectors as in \Fig{level13GFT}. The color indicates the magnitude of the residue, $ |R| < 1$, as shown in the color bar. Orbits with $|R| \ge 1$ are shown as small, black points.}{omegaEps02}{6.5in}

\subsection{Cubic Fields}
KAM theory suggests that the most robust invariant tori should have Diophantine rotation vectors. One way to construct these is to use integral bases of algebraic fields. Recall that $\alpha \in \bC$ is an \emph{algebraic number} of degree $n$ if it is a root of an $n^{th}$-degree polynomial with rational coefficients, but not of any such polynomial of lower degree. The \emph{algebraic field} generated by $\alpha$ is
\[
	\bQ(\alpha) = \left\{ \sum_{i=0}^{n-1} a_i \alpha^{i}: a_i \in \bQ \right\} .
\]
For example, the golden mean $\phi$, the larger root of $\phi^2-\phi-1=0$, is a quadratic irrational and $\bQ(\phi) = \{a + b\phi: a,b \in \bQ\}$.
The ring of integers $\bZ(\alpha) \subset \bQ(\alpha)$ consists of those elements that are roots of monic polynomials, $x^n+p_1 x^{n-1}+\ldots +p_n$, with $p_i \in \bZ$. The ring is an $n$-dimensional vector space over $\bZ$, and a basis $\{\alpha_1,\ldots, \alpha_n\}$ for this space is called an \emph{integral basis}. In general, integral bases give rise to Diophantine rotation vectors.

\begin{thm}[\cite{Cassels57,Cusick72}]\label{thm:Cusick}
If $(\omega,1) \in \bR^{d+1}$ is an integral basis of a real algebraic field of degree $d+1$, then $\omega \in \cD_{d}$.
\end{thm}
%

For example, $(\phi,1)$ is a basis for $\bQ(\phi)$, thus $\bZ(\phi) = \{q\phi+p: q,p \in \bZ\}$. Every integral basis for $\bZ(\phi)$ is of the form $ (\alpha_1,\alpha_2) = (\phi, 1)M$, where $M \in SL(2,\bZ)$. The ratios, $\omega = \alpha_1/\alpha_2$ of these integral bases are precisely the noble numbers, conjectured to correspond to the locally most robust invariant circles of twist maps, recall \Sec{Introduction}.

Thus a natural extension of the robustness conjecture of the nobles for $d=1$ is that the most robust tori for $d=2$ should have rotation vectors in an integral basis of a cubic field. Several candidates have been proposed for the most robust field \cite{Kim86, Lochak92, Tompaidis96a}, and invariant tori with such rotation vectors have been studied for 4D symplectic maps, recall \Sec{Introduction}. However, the robustness conjecture has not yet, to our knowledge, been investigated. Indeed, it would be difficult to find a ``last" invariant torus in the symplectic case: since the tori are not codimension one, they do not form barriers.

\subsection{Spiral Mean}\label{sec:SpiralMean}

One natural generalization of the golden field is that of the \emph{spiral mean}, the real root of 
\beq{SpiralCubic}
	\sigma^3 - \sigma -1 = 0 ,
\eeq
proposed by Kim and Ostlund \cite{Kim86}.
We will denote the spiral mean by $\sigma$ and its conjugates by $\sigma_1$ and $\sigma_2 = \bar{\sigma}_1$; here
\bsplit{SpiralMean}
 \sigma &\equiv \sigma_0 \approx 1.3247179572447460 , \\
 \sigma_{1,2} &= \frac{1}{\sqrt{\sigma}} e^{\pm i\theta} , \quad
 \theta = \arccos(-\tfrac12 \sigma^{\frac32}) \approx 2.437734932288317 .
\esplit 
The spiral mean, like $\phi$, is a ``Pisot-Vijayaraghavan" (PV) number: a root of a monic polynomial with exactly one root outside the unit circle. Indeed, $\phi$ and $\sigma$ are the smallest quadratic and cubic PV numbers, respectively. The discriminant of \Eq{SpiralCubic} is square-free, $\Delta = -23$, so the set $\{\sigma^2,\sigma,1\}$ is an integral basis for $\bZ(\sigma)$.



For our purposes, it is providential that $\bZ(\sigma)$ is naturally related to the generalized Farey tree construction. Indeed the polynomial \Eq{SpiralCubic} is the characteristic polynomial for the cone matrices $C_l$ and $C_r$ of \Eq{GFTMatrices}, and their maximal eigenvectors,
\beq{Eigenvectors}
	v_{\bar l} = \begin{pmatrix} 1 \\ \sigma^2 \\ \sigma\end{pmatrix} , \quad
	v_{\bar r} = \begin{pmatrix} \sigma^2 \\ 1 \\ \sigma\end{pmatrix} ,
\eeq
are in $\bZ(\sigma)$. Moreover, since every entry of the fifth powers of both matrices \Eq{GFTMatrices} is positive, these matrices are irreducible in the Perron-Frobenius sense. As a consequence, for any $v \in \bR_+^3$, $C_i^\ell v \to c\sigma^\ell v_i$ as $\ell \to \infty$. Thus by \Eq{GFTPath}, the directions $v_{\bar l}$ and $v_{\bar r}$ correspond to the paths $p = l^\infty \equiv \bar l$ and $r^\infty \equiv \bar r$, respectively.

As an example, the sequence of integer vectors $(m_\ell,n_\ell)$ with paths $p = r^\ell$ are
\beq{SpiralSequence}
	\vector{1}{1}{0}, \; \vector{1}{0}{1}, \; \vector{1}{1}{1}, \; \vector{2}{1}{1},\; 
	\vector{2}{1}{2}, \; \vector{3}{2}{2}, \; \vector{4}{2}{3}, \ldots, \vector{n_{\ell+1}}{n_{\ell-1}}{n_{\ell}} \ldots .
\eeq
It is not hard to see that the sequence of periods $\{n_0,n_1,\ldots\}$ satisfies the recurrence
\beq{SpiralRec}
	n_{\ell+3} = n_{\ell+1}+n_{\ell}, \quad n_0 = 0,\; n_1 = n_2 = 1 ,
\eeq
which has the solution
\bsplit{SpiralPeriods}
	n_\ell &=a_0\sigma^\ell + a_1\sigma_1^\ell + a_2\sigma_2^\ell , \\
	a_i &= \tfrac{1}{23}(3\sigma_i^2 + 7\sigma_i -2) .
\esplit

The directions of Farey sequence \Eq{SpiralSequence}, $m_\ell/n_\ell = (n_{\ell+1}/n_\ell, n_{\ell-1}/n_\ell)$, limit on the rotation vector
\[
	\omega = (\sigma, 1/\sigma) = (\sigma,\sigma^2-1).
\]
Unlike the equivalent sequence for the golden mean, these successive approximants include rotation vectors that are not ``best" approximants to $\omega$. Recall that a rational $m/n$ is a best approximant to $\omega$ if
\beq{BestApproximant}
	\|n\omega -m\| = \min_{0<q \le n} \min_{p\in \bZ^d} \|q \omega -p\|
\eeq
for some norm $\|\,\|$.
The sequence of best approximants depends upon the choice of norm and, unlike the classical Farey tree, it is not known if the generalized Farey tree gives all of the best approximants. However, up to period $50,000$, the best approximants to $v_{\bar r}$ in the $1$, $2$ and $\infty$ norms are the same; they have the periods
\[
	n_\ell \in \{ 1, 3, 9, 12, 37, 114, 151, 465, 1432, 1897, 5482, 17991, 23833\}.
\]
Moreover, we found that all of the best approximants for these three norms can be found in the Farey sequence \Eq{SpiralSequence} up to at least period $10^6$.

\section{Residue Criterion for a Spiral Mean Torus}\label{sec:ResidueSpiralMean}
An invariant torus on which the dynamics is diffeomorphic to rigid rotation must have each Lyapunov multiplier corresponding to its tangent directions equal to one. Consequently, all of the multipliers of a codimension-one torus in a volume-preserving map must be one. Thus a natural conjecture is that the multipliers of a sequence of periodic orbits that limit on such a torus should all limit to one. Similar considerations for the symplectic case led to the residue bounds discussed in \Sec{ResidueCriterion}. As a consequence, for the three-dimensional volume-preserving case, periodic orbits in the neighborhood of a smooth invariant torus should have stability parameters near the point $\tau_1=\tau_2=3$ that corresponds to a triple-one multiplier, recall \Fig{Stability}. 

When the map is reversible and the reversor fixed sets are graphs over the action (as in \Eq{FixedSets}), then every rotational invariant torus will contain symmetric orbits. If the torus has an incommensurate rotation vector, then these orbits will be dense. Thus it seems sensible to investigate the existence of such tori by looking for sequences of symmetric periodic orbits whose residues, \Eq{Residue}, approach zero. 

\subsection{A Spiral Torus}
In this section we explore the generalization of Greene's residue criterion for one particular spiral mean torus. Here we study the direction
\beq{LLR}
	ll\bar r \simeq (1, \sigma^3, \sigma^4)^T .
\eeq
Projecting this direction onto the plane $(\omega,1)$ gives the rotation vector
\beq{SpiralOmega}
	\omega = (\sigma-1,\sigma^2-1) \in [0,1]\times [0,1].
\eeq
The Farey convergents for the path \Eq{LLR} are
\[
	llr^\ell \simeq (n_{\ell-1},n_{\ell+2},n_{\ell+3})^T
\]
where $n_\ell$ obeys \Eq{SpiralRec}.
Several of the vectors in this sequence are shown in \Tbl{llrrrSeq}


\begin{table}[htbp]
 \centering
 \begin{tabular}{@{} lc|lc @{}} 
 	path & $(m,n)$ & path & $(m,n)$\\
  \hline
 	$ll$ &$(1,1,1)$ &		$llr^{20}$&$(114,265,351)$\\
	$llr$ & $(0,1,2)$ &		$llr^{21}$&$(151,351,465)$\\
	$llr^2$ & $(1,2,2)$ &		$llr^{22}$& $(200,465,616)$\\
	\multicolumn{2}{c|}{\vdots} & \multicolumn{2}{c}{\vdots} \\
	$llr^6$ & $(2,5,7)$ &	$llr^{33}$ & $(4410,10252,13581)$\\
	$llr^7$ & $(3,7,9)$ &	$llr^{34}$ & $(5842,13581,17991)$\\
	$llr^8$ & $(4,9,12)$ &	$llr^{35}$ & $(7739,17991,23833)$\\
	$llr^9$ & $(5,12,16)$ &	$llr^{36}$ & $(10252,23833,31572)$\\
	\multicolumn{2}{c|}{\vdots} & \multicolumn{2}{c}{\vdots} \\
 \end{tabular}
 \caption{\footnotesize Integer vectors for the generalized Farey paths $llr^\ell$.}
 \label{tbl:llrrrSeq}
\end{table}

When $\eps = 0$, the $llr^\ell$ periodic orbits are, by \Eq{ZDeltaStar}, at 
\beq{SpiralZEps0}
	z_\ell=\frac{n_{\ell-1}}{n_{\ell+3}} -\gamma , \quad
	\delta_\ell = \beta z_\ell^2 - \frac{n_{\ell+2}}{n_{\ell+3}} .
\eeq
These values converge geometrically to the position of the spiral torus at $\eps = 0$
\begin{align*}
	z_\infty &= \sigma-1-\gamma \approx -0.2933160310 ,\\
 \delta_\infty &= \beta z_\infty^2 +1-\sigma^2 \approx -0.5828090783 ,
\end{align*}
as can be seen in \Fig{SMzconv}(a).
Indeed, from \Eq{SpiralPeriods} $n_\ell \sim a_0\sigma^\ell+ \cO(\sigma^{-\ell/2})$, so the convergents approach this point geometrically at a rate $\sigma^{-3/2}$; for example,
\beq{AsymptoticZ}
	z_\ell 
	  \approx z_\infty + \frac{0.6037}{\sigma^{3\ell/2}} \cos(\ell\theta +0.4892)
			+ \cO( \sigma^{-3\ell} ) 
\eeq
with $\theta$ from \Eq{SpiralMean}. 
The line in \Fig{SMzconv}(a) corresponds to \Eq{AsymptoticZ} with the cosine replaced by $1$; it is an effective upper bound to the values \Eq{SpiralZEps0}. The modulation of this geometrical convergence by the phase $\theta$ is illustrated by the (green) curve in the figure, which is obtained from \Eq{SpiralZEps0} by treating $\ell$ as a continuous variable. The near periodicities of the discrete levels can be explained by the fact that $\theta \approx \frac{4\pi}{5}$ and even more closely $\theta \approx \frac{10\pi}{13}$.

\InsertFigTwo{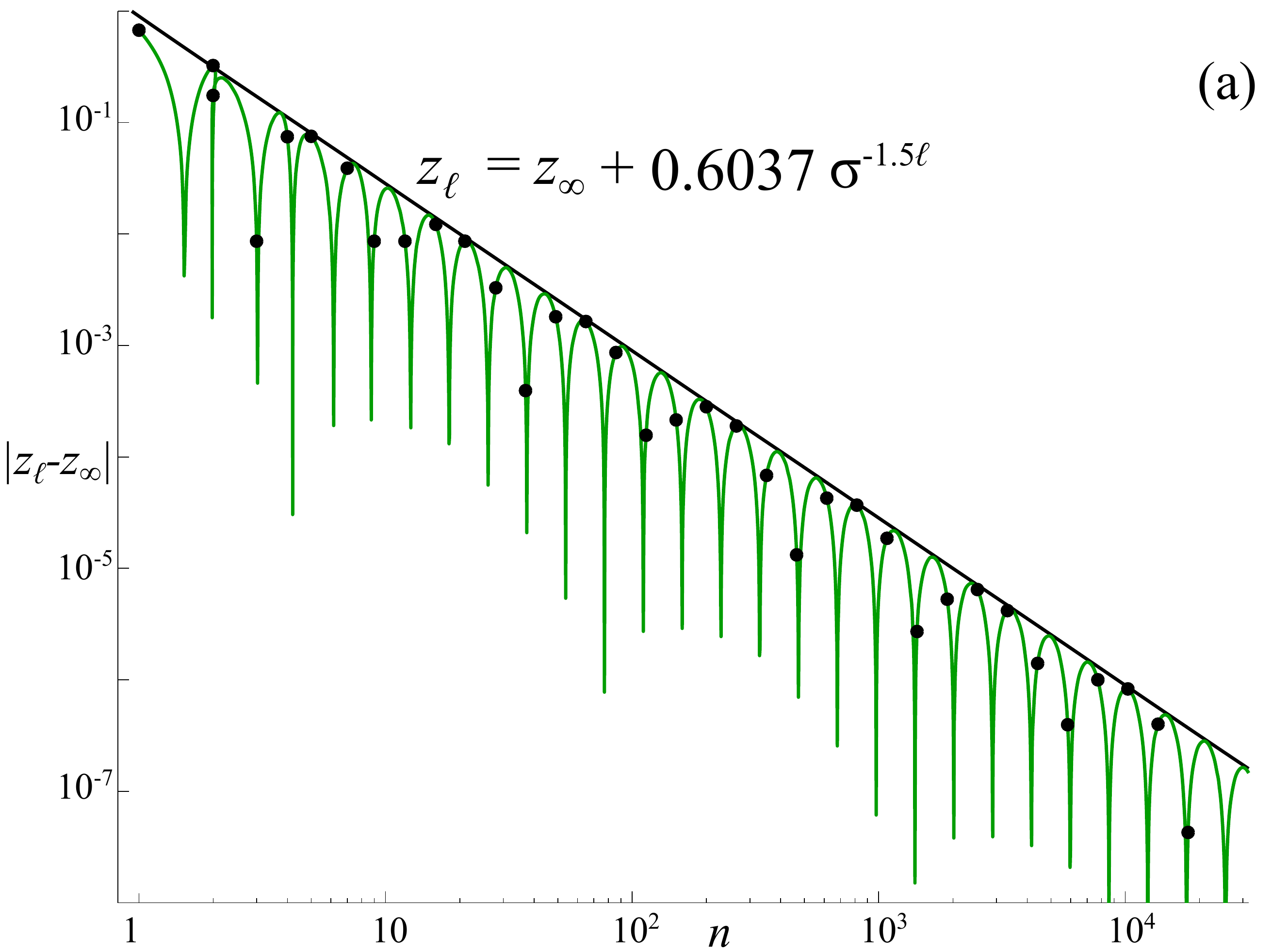}{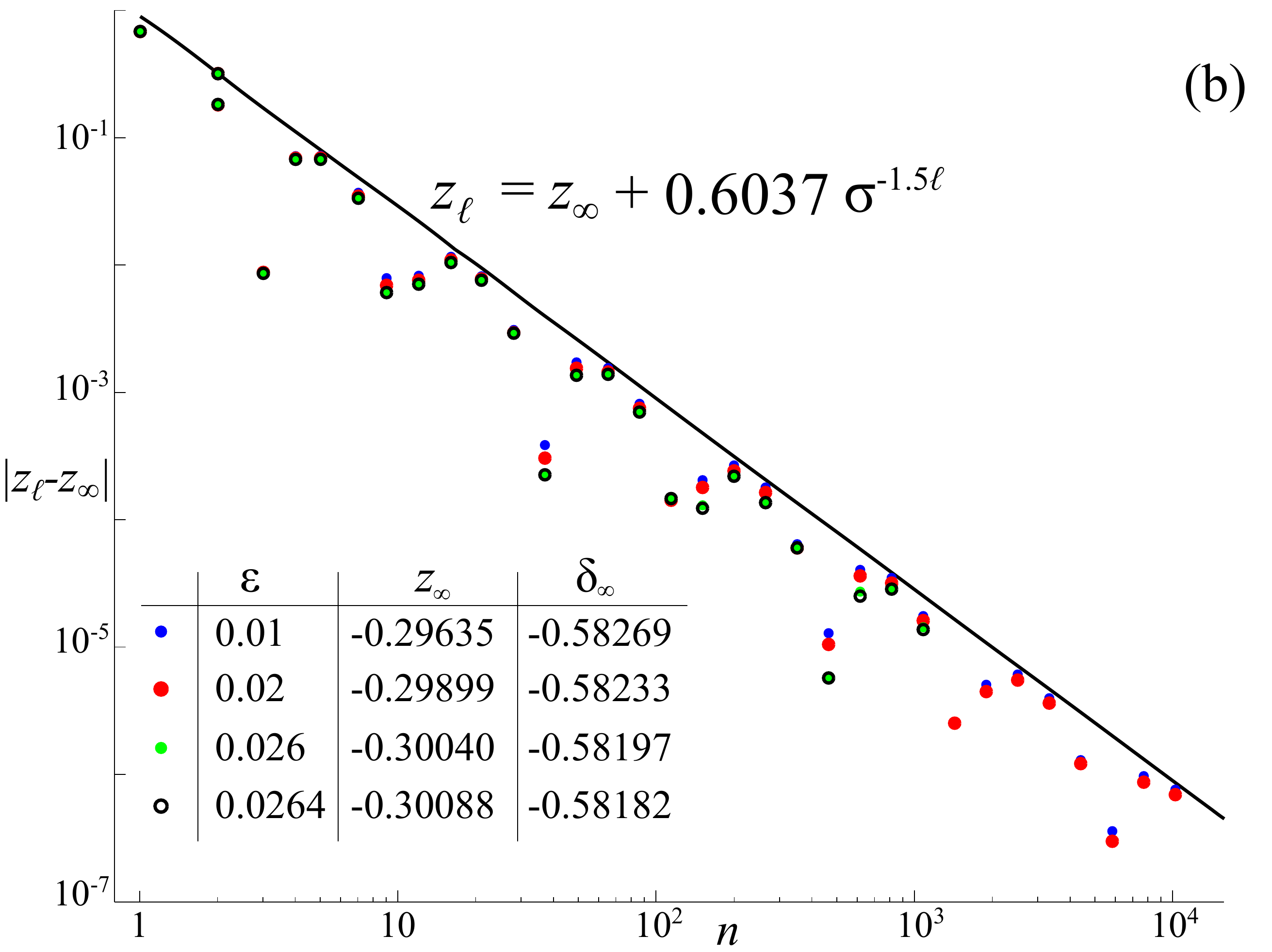}{The initial action for the first 35 approximants to $llr^\infty$ for (a) $\eps=0$ and (b) four cases with $\eps > 0$. The curve (green) in panel (a) is the exact result \Eq{SpiralZEps0} upon extending the level $\ell$ to a continuous variable. The line (black) in both panels is the upper bound of the asymptotic result \Eq{AsymptoticZ} with slope $-\frac32$ on the log-log scale.}{SMzconv}{3.4in}

It is interesting that the geometric convergence of $z_\ell$ appears to be maintained for $\eps >0$---computations of the actions of orbits for the spiral convergents for four values of $\eps$ are shown in \Fig{SMzconv}(b). In this figure, data is not shown for periods larger than $10^3$ for the two largest values of $\eps$ ($0.026$ and $0.0264$) because the orbit finding algorithm did not converge. Note that even though, as shown in the inset, the converged point $(z_\infty,\delta_\infty)$ varies with $\eps$, the geometric convergence rate is consistent with the $\eps = 0$ value $\sigma^{-3/2}$; however, there is some indication that the convergence rate could be faster for the larger two $\eps$'s. Unfortunately, the oscillations make it difficult to extract a precise rate. As we will see below, for $\eps > 0.0259$ the torus seems to be destroyed, and by analogy with the standard map \cite{MacKay84}, a different convergence rate might be expected.

\subsection{Critical Torus}\label{sec:CriticalTorus}

Recall from \Fig{ResidueError}(b) that the residues of the level-$\ell$ orbits appear to converge to zero for fixed, small $\eps$, but from \Fig{AsymptoticResidue} that each grows geometrically with $\eps$. As can be seen in \Fig{SMResidues}(a), the residues of low period spiral mean approximants begin to grow near $\eps \approx 0.03$, though for periods larger than $23$, the onset of rapid growth in $R$ occurs near $\eps = 0.26$. For larger period spiral approximants, as shown on a log scale in \Fig{SMResidues}(b), rapid growth in $R$ begins near $\eps \approx 0.0259$. This evidence suggests that the $llr^\infty$ torus does not exist for $\eps > 0.0259$.

\InsertFigTwo{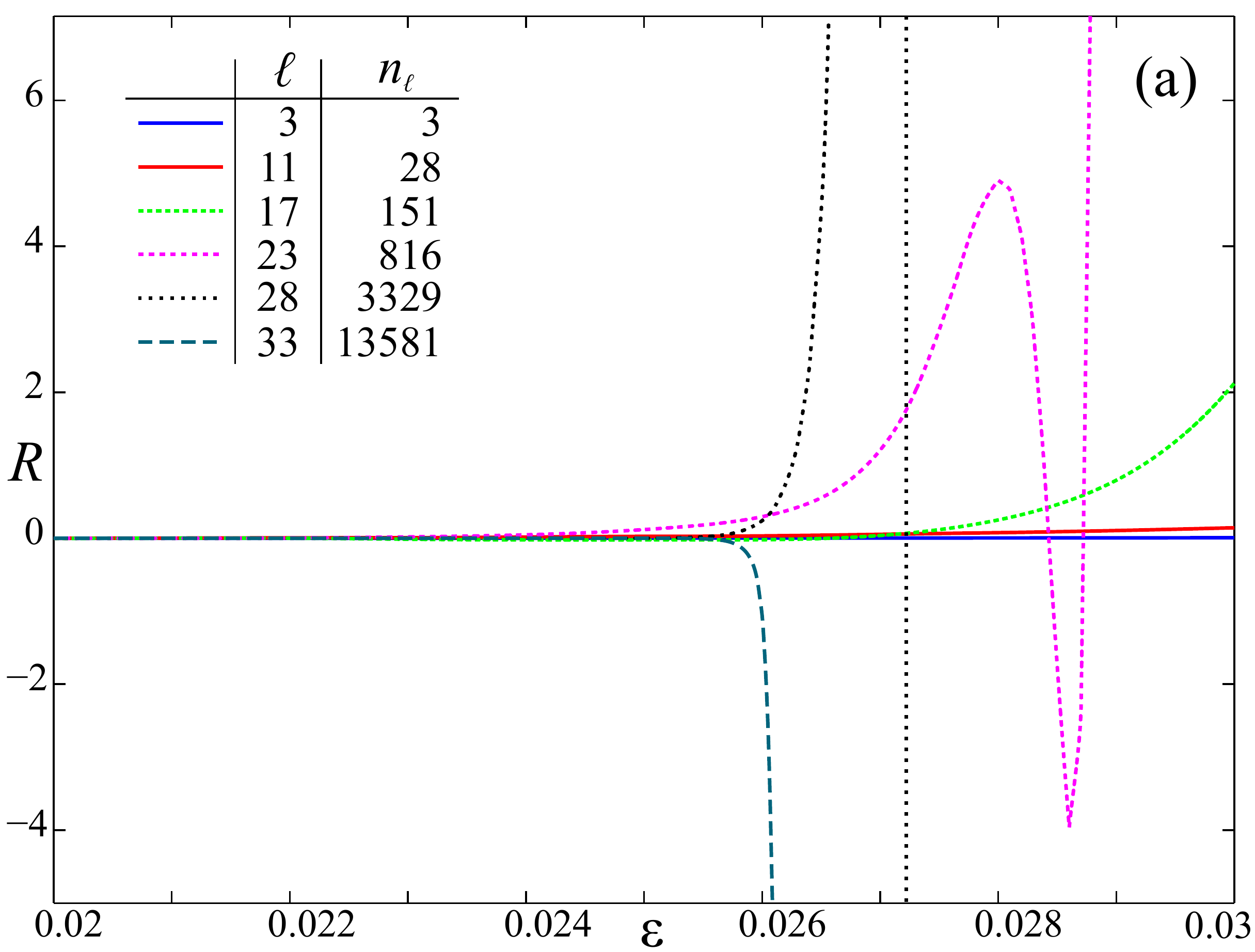}{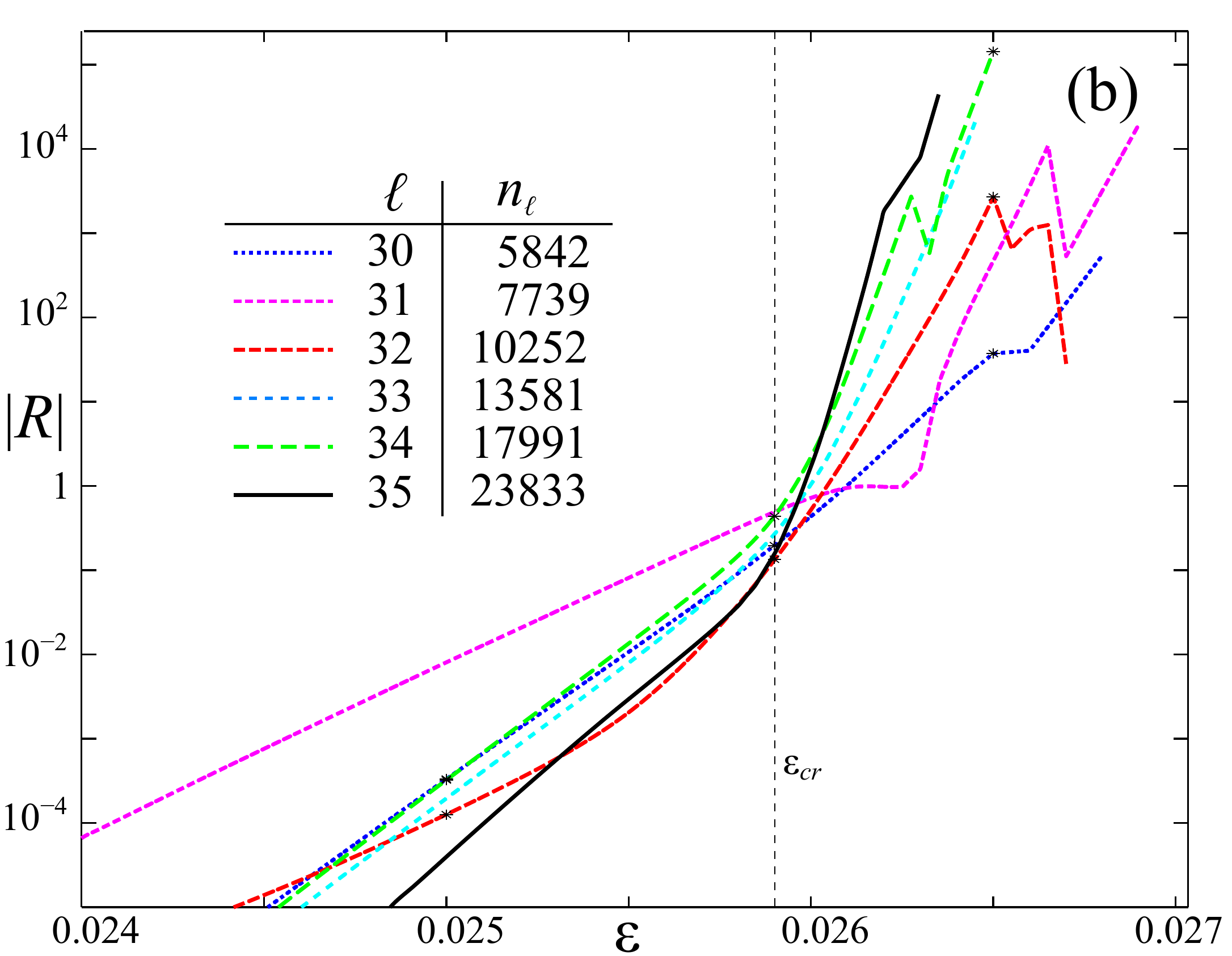}{Residue as a function of $\eps$ for $\fix{S_1}$ orbits with rotation vectors to $llr^\ell$ for (a) six low-period orbits, and (b) six longer period orbits near $\eps_{cr}$. The stars indicate the residues for the three values of $\eps$ shown in \Fig{SpiralTori}.}{SMResidues}{3.4in}

To more systematically investigate the residue criterion, we choose a threshold residue, $R_{th}$, and compute the smallest positive $\eps$ for which the level $\ell$ orbit has $|R| = R_{th}$; call this value $\eps_{R_{th}}$. Computationally, we continue each orbit in $\eps$, as discussed in \Sec{Numerics}, until $R_{th}$ is bracketed. Bisection is then used to find $\eps_{R_{th}}$ as precisely as possible. The threshold $\eps$ values are shown for the first 36 levels, up to period $31572$, in \Tbl{SpiralMeanEpsCR}. These thresholds appear to converge to
\beq{SprialEpsCr}
	\eps_{cr} \approx 0.02590 \pm 5(10)^{-5}, 
\eeq
independently of $R_{th}$. Computations for the spiral mean orbits on the other symmetry lines gave the same result.

\begin{table}
\centering 
\begin{tabular}{r|r l l l ||r|rlll}
$\ell$ &$n_\ell$ & \multicolumn{1}{c}{$\eps_{0.5}$} & \multicolumn{1}{c}{$\eps_{0.9}$} & \multicolumn{1}{c|| }{$\eps_{1.5}$} & 
$\ell$ &$n_\ell$ & \multicolumn{1}{c}{$\eps_{0.5}$} & \multicolumn{1}{c}{$\eps_{0.9}$} & \multicolumn{1}{c}{$\eps_{1.5}$}\\
\hline
0& *1 &0.159156 & 0.286479&0.477464 & 19 & 265 &0.028418 &0.028739&0.029030 \\ 
1& 2 &0.079578 & 0.143239&0.238734 & 20 & 351 &0.028425 &0.028532 &0.028650\\ 
2& 2 &0.121723 & 0.152553&0.188191 & 21 & *465 &0.027723 &0.028382&0.028444\\ 
3& *3 &0.097264 & 0.116817&0.137303 & 22 & 616 &0.026147 &0.026563&0.026920\\ 
4& 4 &0.047052 & 0.068028&0.154364 & 23 & 816 &0.026426 &0.026828&0.027130\\ 
5& 5 &0.038700 & 0.053493&0.135967 & 24 & 1081 &0.026372 &0.026551&0.026696\\ 
6& 7 &0.094469 & 0.097173&0.100279 & 25 & *1432&0.026746 &0.026791&0.026840\\ 
7& *9 &0.032182 & 0.042449&0.068823 & 26 & *1897&0.026046 &0.026209&0.026341\\ 
8& *12 &0.035600 & 0.040047&0.044786 & 27 & 2513&0.025880 &0.026076&0.026218\\ 
9& 16 &0.038450 & 0.041201&0.043517 &28 &3329&0.026136 &0.026240&0.026327\\ 
10& 21 &0.030190 & 0.032730&0.035036 & 29 & 4410 &0.026058 &0.026169&0.026262\\ 
11& 28 &0.034410 & 0.036471&0.038134 & 30 & *5842&0.026016 &0.026085&0.026143\\ 
12& *37 &0.029858 &0.032166&0.038226 & 31 & 7739&0.025905 &0.026073&0.026316\\ 
13& 49 &0.030033 &0.031782&0.033172 & 32 & 10252&0.025998 &0.026037&0.026072\\ 
14& 65 &0.028867 &0.031821&0.033347 & 33 &13581&0.025949 &0.025990&0.0260 \\ 
15& 86 &0.028891 & 0.029774&0.030563 & 34 &*17991&0.02591 &0.02595&0.0260\\ 
16& *114 & 0.029003& 0.029824&0.030530 & 35 &*23833&0.02595 &0.025976&0.0260\\ 
17& *151 & 0.028573 & 0.029119&0.029631& 36 &31572&0.0258 &0.0259 &0.0259\\ 
18& 200&0.026830 &0.028130 &0.030583 & & & & &\\ 
\\ 
\end{tabular}
\caption{\footnotesize Stability thresholds for the first $36$ spiral mean approximants of $llr^\infty$ on $\fix{S_1}$. Shown are the values $\eps_{R_{th}}$ at which the residue first reaches three thresholds, $R_{th} = 0.5$, $0.9$, and $1.5$. The * denotes orbits that are best approximants to $\omega$ in the sense of \Eq{BestApproximant}.}
\label{tbl:SpiralMeanEpsCR}
\end{table}

The convergence of the thresholds to $\eps_{cr}$ with period is geometric, as shown in \Fig{SMeps_cr} for $R_{th} = 0.9$. A least squares fit to this data gives
\beq{fit}
	\eps_{\ell}=\eps_{cr} + (0.17 \pm 0.06)n^{-0.79 \pm 0.02} .
\eeq
It is perhaps interesting that the rate of convergence, $0.79$, is close to, but significantly different from, $\sigma^{-1} \approx 0.75488$.

\InsertFig{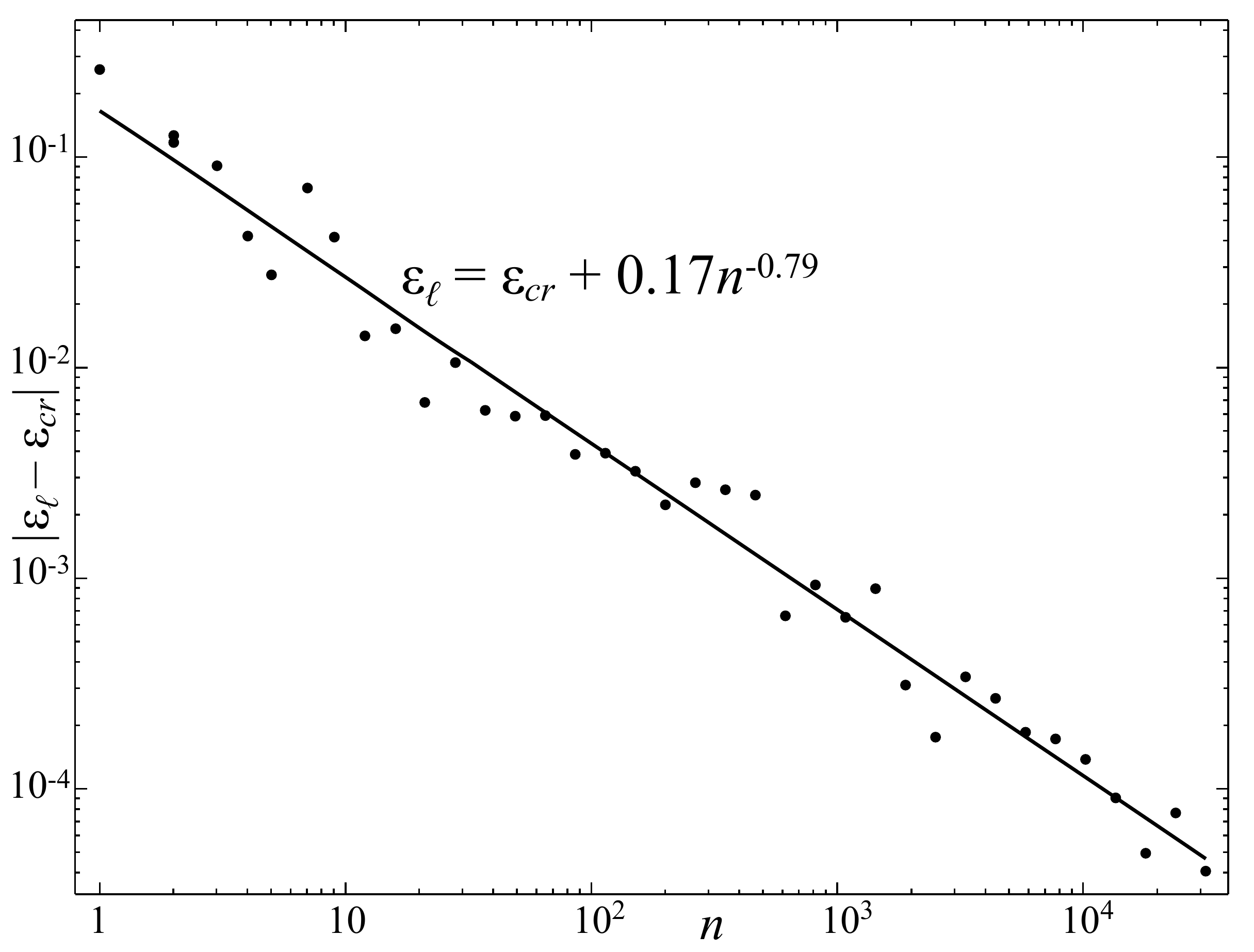}{The $\eps$ values for $R_{th}=0.9$ for the first 36 approximants to $llr^\infty$ using $\eps_{cr} = 0.02590$.}{SMeps_cr}{3.4in}

\subsection{Supercritical Tori}\label{sec:Destruction}
These results imply that the $llr^\infty$ torus appears to be critical at $\eps_{cr} \approx 0.02590$. The limit of the $\fix{S_1}$ periodic orbits at this value is
\[
	(x_\infty, z_\infty, \delta_\infty) \approx (0, -0.300443, -0.581962) ,
\]
which appears to lie on on a critical spiral mean torus.

By analogy with the area-preserving case, we expect that when $\eps > \eps_{cr}$, the $llr^\ell$ sequence will still converge to a quasiperiodic orbit, but one that is not dense on a torus; indeed, for twist maps the supercritical orbits are dense on Cantor sets. Figure \ref{fig:SpiralTori} gives a visualization of this for our map: it shows the phase portraits of three spiral approximants for three values of $\eps$ near $\eps_{cr}$. The residues for these nine cases were marked with a $*$ in \Fig{SMResidues}(b). When $\eps < \eps_{cr}$ (the left column of \Fig{SpiralTori}) the approximating orbits are stable and as the period is increased the points appear to limit to a smooth, nonzero density on a smooth torus. In the middle column of the figure, where $\eps = \eps_{cr}$, the density becomes less smooth, and in the right column, where $\eps > \eps_{cr}$, it appears to limit to a density with holes. This can be better seen in \Fig{Spiral1779} which shows the projection of the $\ell = 34$ approximant onto the $(x,y)$ plane.

\InsertFig{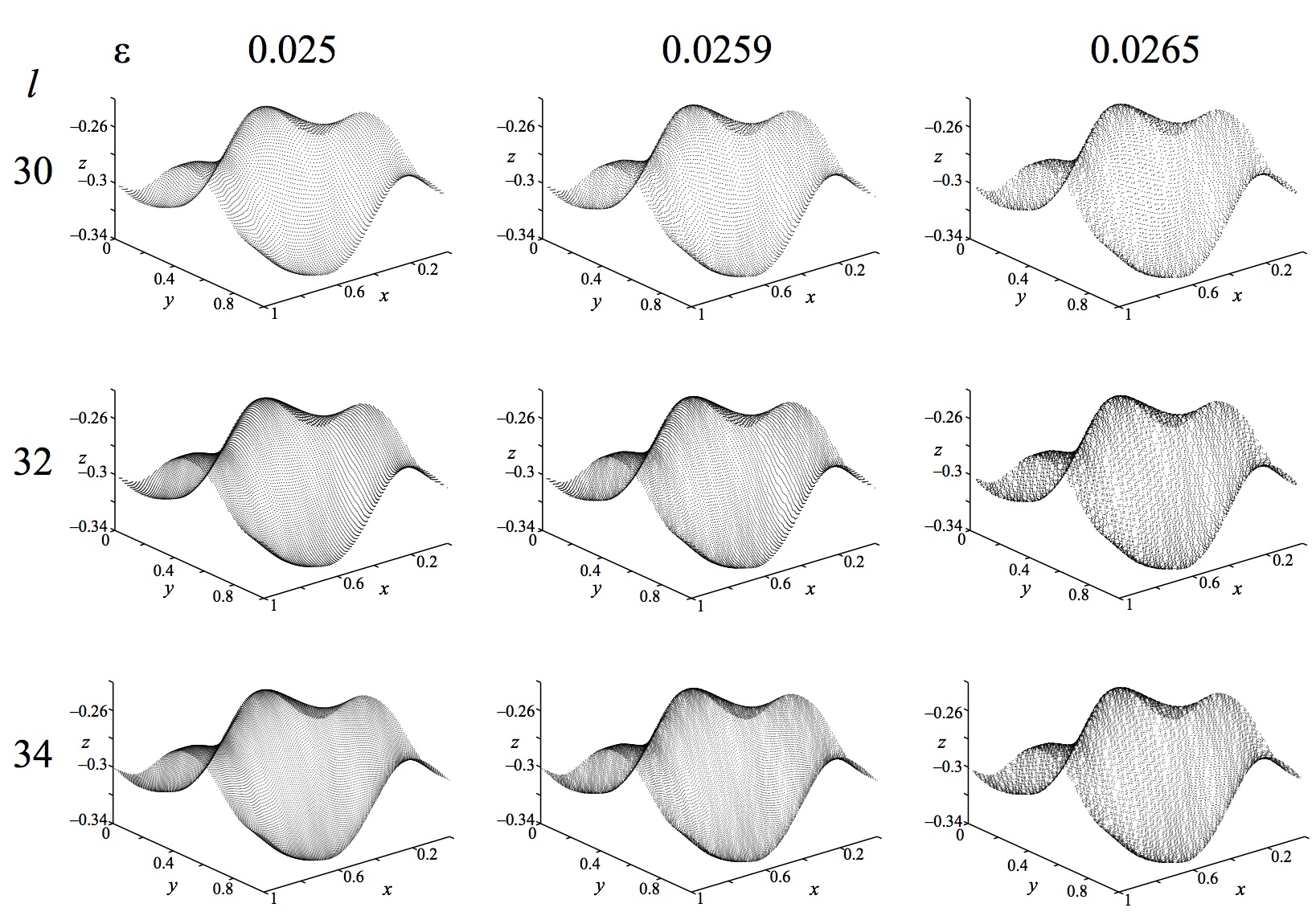}{Phase portraits of the periodic orbits $llr^{30}$, $llr^{32}$, and 
$llr^{34}$ (rows) on $\fix{S_1}$ for three $\eps$ values (columns). For $\eps < \eps_{cr} 
\approx 0.0259$ the orbits appear to limit on a smooth torus. At $\eps_{cr}$ the density is nonuniform, and when $\eps > \eps_{cr}$ the orbits seem to limit to a torus with holes.}{SpiralTori}{7in}

\InsertFig{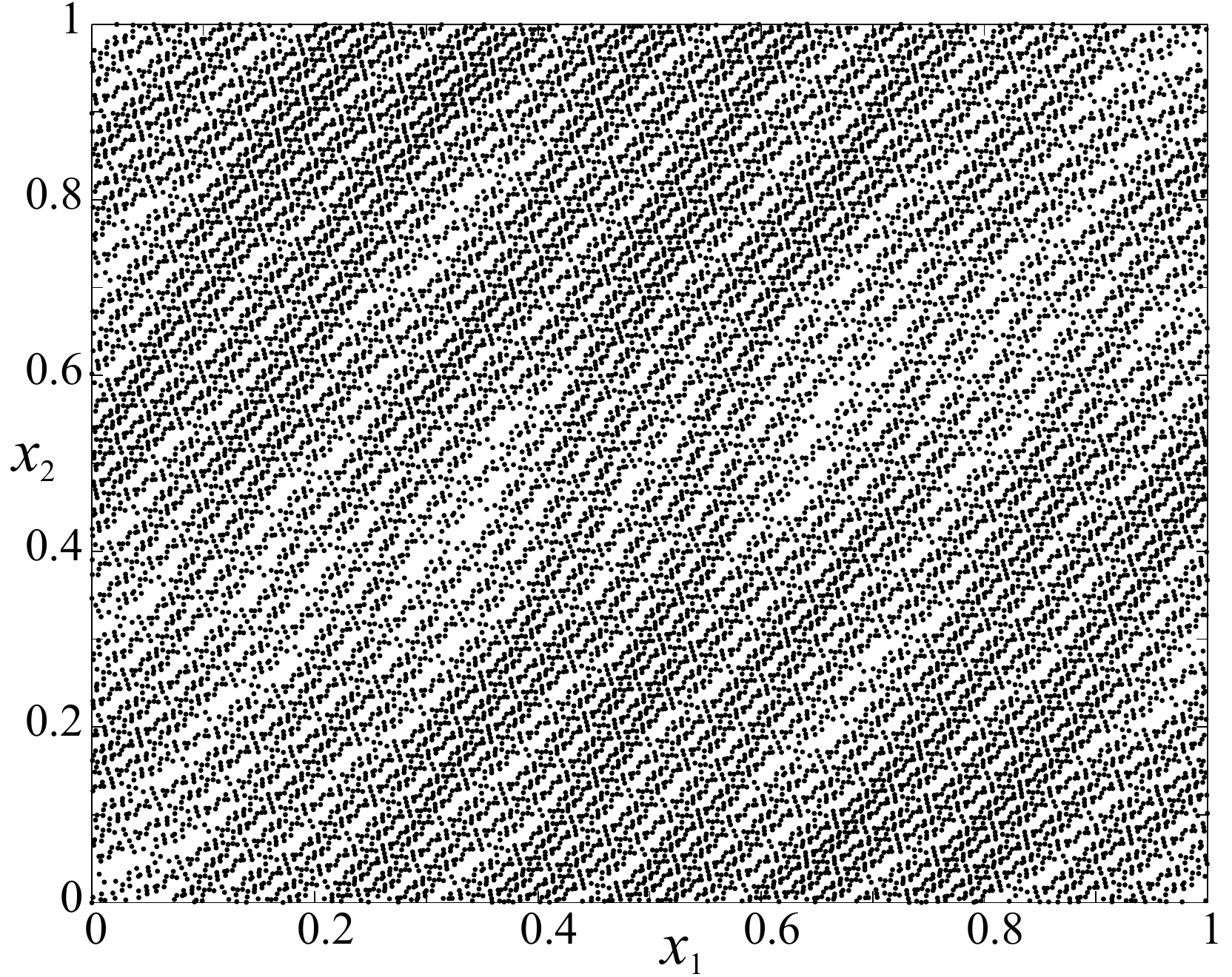}{Projection of the $\fix{S_1}$ spiral approximant $llr^{34}$, with $(m,n) = (5842, 13581, 17991)$, onto the angle-plane for $\eps= 0.0265$.}{Spiral1779}{3.4in}

\subsection{Varying Parameters}
Up to this point we have studied only the standard parameter set \Eq{StdParams}; in this section we vary the parameter $c$ in the force \Eq{Force}. Note that the map \Eq{VPMap} is invariant under the replacement $(x,\eps,c) \to (x+(\tfrac12,\tfrac12),-\eps,-c)$. Since we have observed that $\eps_{cr}$ is the same for the $\fix{S_1}$ and the $\fix{S_1 \circ T_{1,1}}$ orbits, the critical graph $(\eps_{cr},c_{cr})$ will be symmetric under reflection through the origin. Thus we study only positive values of $\eps$.

Figure \ref{fig:ChangeC} shows the critical set $(\eps_{cr},c_{cr})$ of the spiral torus \Eq{LLR} for $c \in [-2,2]$, holding the remaining parameters fixed according to \Eq{StdParams}. On this scale, $\eps_{cr}$ appears to be a graph over $c$, and for $c< 0$ this graph appears to grow smoothly as $c$ increases. The peak value, $\eps_{cr} \approx 0.06156$, occurs at $c_{cr} \approx 0.07281$. Though the graph appears to decrease monotonically for larger $c$ values, this is not the case: there are a number of cusps. These appear to be associated with an increasing number of oscillations in the function $R(\eps)$ before it settles into its asymptotic growth, recall \Sec{NumericalResidues}. 
An enlargement near the most prominent cusp is shown in \Fig{ChangeC}(b). 

The oscillations in $R(\eps)$ make it harder to obtain accurate values for the critical parameters near each cusp. To resolve the cusp shown in \Fig{ChangeC}(b), we used a longer period orbit (period $31572$), and a larger threshold $R_{th} = 10$. Note that what appeared to be a single peak near $c=0.073$ becomes, in the enlargement, a number of cusps and several possible discontinuities. Each cusp corresponds to a change of the large $\eps$ asymptotic behavior of $R$, and near the cusp the number of oscillations in the graph $R(\eps)$ changes. We are not able to resolve the apparent discontinuities in \Fig{ChangeC}(b)---it is possible that the critical set is not a graph over $c$ in these regions.
Several less prominent cusps are also visible in \Fig{ChangeC}(a); the local behavior near these is similar to the one shown in the enlargement.

\InsertFigTwo{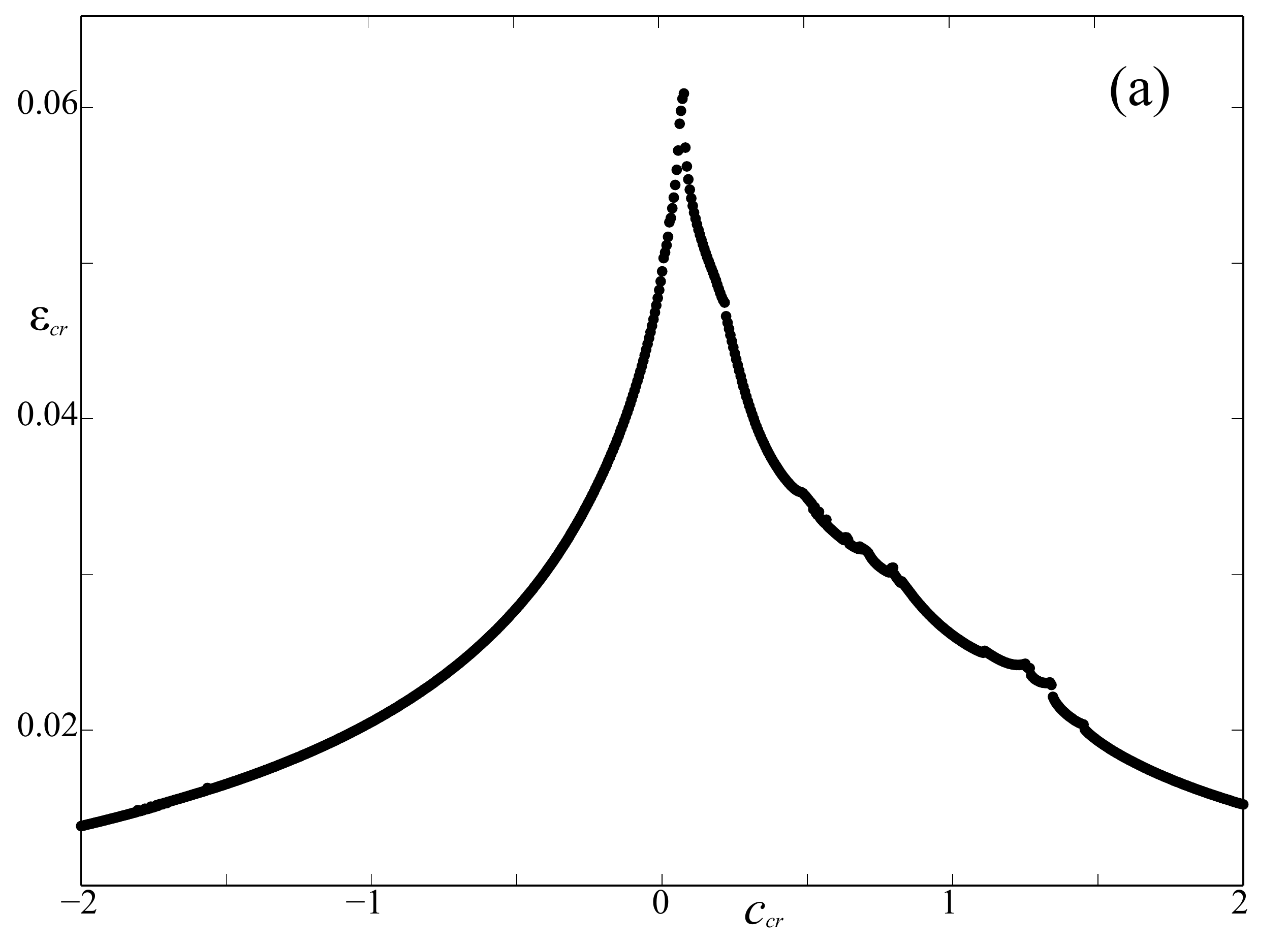}{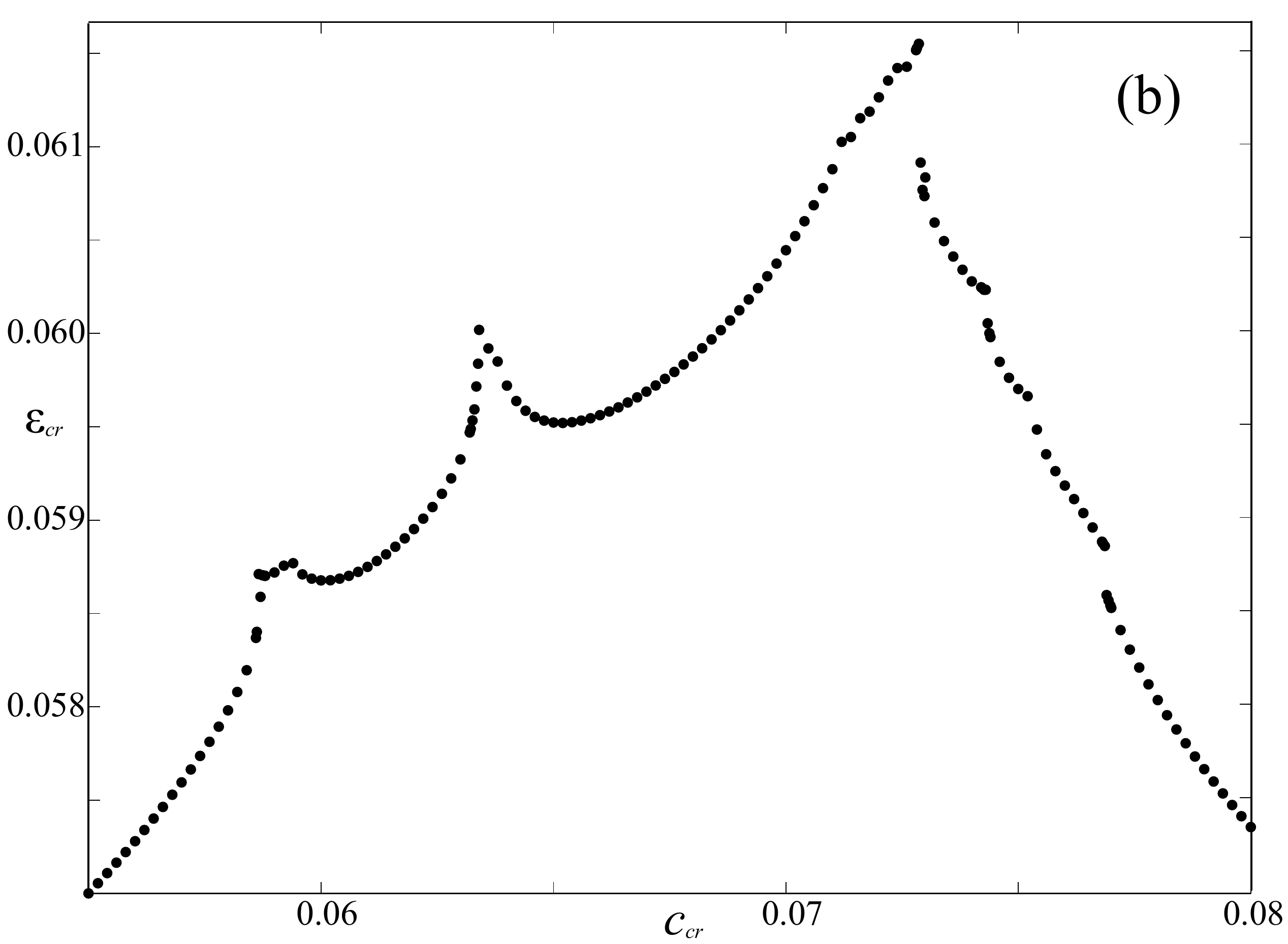}{Critical set $(\eps_{cr},c_{cr})$ for the $llr^\infty$ torus with the remaining parameters fixed as usual at \Eq{StdParams}. (a) Critical set computed using the $llr^{30}$, period $5482$, orbit with $R_{th}=0.9$. (b) Enlargement near  the peak $\eps_{cr}$ using the $llr^{36}$ orbit with $R_{th} = 10$.}{ChangeC}{3.4in}

\section{The Last Torus}\label{sec:LastTorus}
The critical function, $\eps_{cr}(\omega)$ is the value of $\eps$ at which the torus with rotation vector $\omega$ is first destroyed upon perturbation from integrability, $\eps = 0$ (here we assume that $\delta= \left< z^2\right> - \omega_2$, recall \Eq{DeltaEq}). When $\omega$ is resonant, $p \cdot \omega = q$, the torus is typically destroyed for any $\eps \neq 0$; i.e., $\eps_{cr}(\omega) = 0$ at resonance. However, KAM theory implies that $\eps_{cr}(\omega) > 0$ when $\omega$ is Diophantine; thus, the critical function is highly singular.

Our goal in this section is to compute $\eps_{cr}(\omega)$ and to identify the most robust torus of our model \Eq{VPMap}. Ideally we would like to discover the analogue of the golden mean, which, as Greene conjectured, appears to give the most robust invariant circle of the standard map \Eq{StdMap}. Though there is considerable numerical support for Greene's conjecture, it has never been proven.\footnote
{It is known that the standard map has no rotational circles when $2\pi \eps > 63/64$ \cite{MacKay85}.}
Strong numerical evidence also indicates that invariant circles with ``noble" rotation numbers,
\[
	\omega = \frac{a\phi+b}{c\phi+d}\;, \mbox{ with } ad-bc = \pm 1,
\]
are \emph{locally} robust for area-preserving maps \cite{Percival82, MacKay92c}. That is, in any interval of rotation numbers the most persistent invariant circle has a noble rotation number. Recall that each noble number is the projection of the integral basis $(a\phi+b,c\phi+d)$ of $\bZ(\phi)$ onto $(\omega,1)$.

\subsection{Critical Function}\label{sec:CritialFunction}
Thus it is natural to conjecture that the most robust tori for a two-angle map may correspond to integral bases of some cubic field. As a preliminary investigation, we will compute the critical function for bases of $\bZ(\sigma)$.
Recall that $v_{\bar r}$ \Eq{Eigenvectors} is one such integral basis and that every other integral basis is obtained from $v_{\bar r}$ by multiplication with some element of $SL(3,\bZ)$. To construct a sampling of these matrices we will use the $2^\ell$ directions on the generalized Farey tree at level $\ell$, recall \Sec{KimOstlund}. To each path $h = i_1i_2\ldots i_\ell$ with $i_j \in\{r,l\}$ there is a corresponding matrix $C_{i_1}C_{i_2}\ldots C_{i_\ell} \in SL(3,\bZ)$ using \Eq{GFTMatrices}. Therefore, the direction 
\[
	v_p = C_{i_1}C_{i_2}\ldots C_{i_\ell} v_{\bar r} \;, \quad i_j \in \{l,r\},
\]
is an integral basis for $\bZ(\sigma)$ with path $p = h\bar r$. We refer to $h$ as the \emph{head} and $\bar r$ as the \emph{tail} of this path. As we remarked in \Sec{KimOstlund} some of the finite paths at level $\ell$ on the generalized Farey tree correspond to duplicated directions; however, the $2^\ell$ directions obtained by appending the infinite tail $\bar r$ are unique.

The projection of each $v_p$ onto $(\omega_p,1)$ gives a rotation vector $\omega_p$ whose components are Diophantine by \Th{Cusick}. Since the matrices $C_l$ and $C_r$ do not generate $SL(3,\bZ)$, the Farey tree does not generate all of the integral bases. 
Nevertheless, as was seen in \Fig{level13GFT}, the resulting rotation vectors $\omega_p$ give a discrete sampling of frequency space that is sensibly more dense in the nonresonant 
regions.

We estimate $\eps_{cr}(\omega_p)$ for each such invariant torus using the residue criterion for nearby periodic orbits. To give a reasonable approximation, but limit the computation time, we approximate the torus using the periodic orbit with path $hr^k$, selecting the length, $k$, of the tail so that the orbit has period larger than $5000$. Note that the $k$ we choose depends on the head $h$. We then compute the value of $\eps$ for which $|R| = R_{th} = 0.9$ for each of the corresponding periodic orbits $hr^k$. 

\InsertFig{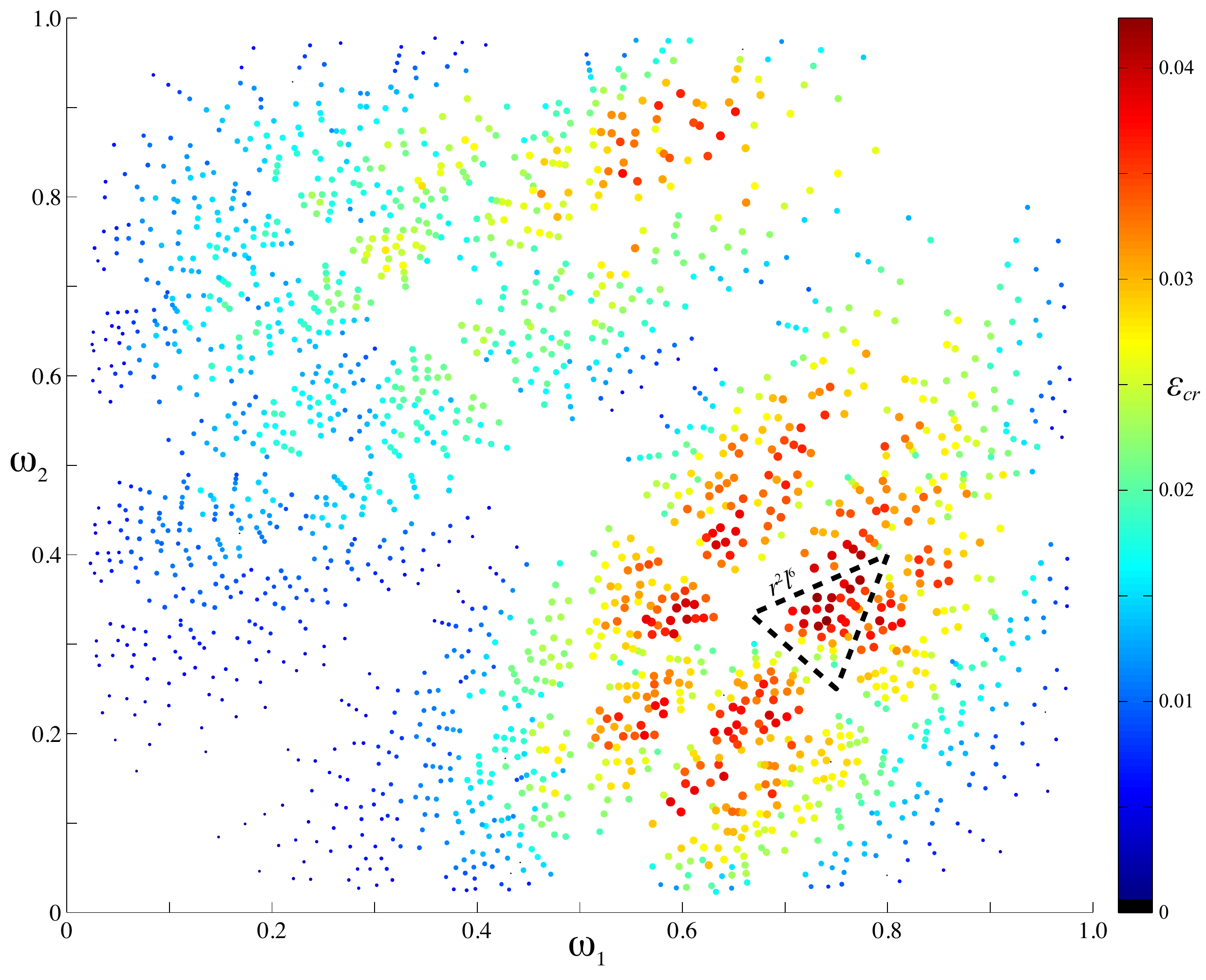}{The critical function for \Eq{VPMap} with the standard parameters \Eq{StdParams}. The values of $\eps_{cr}$ are low near the resonances, but grow into several peaks. The level-$8$ triangle that contains the most robust torus, $r^2l^6$, is shown (dashed). }{cfun}{5in}

To sample the tori with $\omega \in [0,1]^2$, we began by generating all of the level $\ell = 13$ heads that start with $rrl$ or $llr$, namely those in the two triangles of \Fig{GFT}(b). The result, shown in \Fig{cfun}, is an approximation to $\eps_{cr}(\omega)$ on the $2 \times 2^{10} = 2048$ rotation vectors at this level. In this figure, the size of the dots (and their color, as shown in the color bar) is proportional to $\eps_{cr}(\omega_p)$. Note that there are several local peaks in the critical function, and that it tends to be small near low-order resonance lines. The most robust torus in this figure has the path $p = r^2 l^6 r^3 l^2 r^k$ with $k=19$; this corresponds to $(m,n) = (3901,1718,5268)$, or rotation vector $\omega_p \approx (0.7405,0.3261)$. For this torus, $\eps_{cr}(\omega_p) = 0.0424$.

Hypothesizing that the most robust torus is nearby, we zoom-in by focusing on the level-$8$ triangle
\[
	r^2l^6 \simeq \begin{pmatrix} 2 & 3 & 4\\
							1 & 1 & 2 \\
							3 & 4 & 5
			 \end{pmatrix}
\]
that contains this vector, i.e., the dashed triangle in \Fig{cfun}.  We now generate the $2^9$, level-$17$ paths within this triangle, i.e., each path of length $17$ that begins with $r^2l^6$. Appending $r^k$ to each of these to give a periodic orbit with period larger than $5000$, gives the zoomed-in view of the critical function shown in \Fig{cfun23}(a). The most robust torus of the $512$ sampled
vectors has $\eps_{cr}(\omega_p) = 0.0479$. It has the path $p = r^2l^6 (rlr)^2 lr^2 r^{k}$ with the $k=17$ approximation $(m,n) = (4276, 2081, 5806)$.  We again zoom-in near the most robust torus, incrementing the level of the base triangle by three, and focusing now on the level-$11$ triangle
\[
	r^2l^6rlr \simeq \begin{pmatrix} 6&2&7\\ 
								3&1&3 \\
								8&3&10
				\end{pmatrix} .
\]
The critical function for another $2^9$ rotation vectors using this level-$11$ triangle as the base is shown in \Fig{cfun23}(b). The most robust rotation vector at this stage has the path $p = r^2l^6 (rlr)^2 l^5 r r^k$ with its $k=14$ approximation  $(m,n) = (4670,2278,6353)$ 
and $\eps_{cr}(\omega_p) = 0.0485$.
This vector is contained in the level-$14$ triangle
\[
	 r^2l^6(rlr)^2 \simeq \begin{pmatrix} 13&6&14\\
	 								6&3&7\\
									18&8&19
					\end{pmatrix},
\]
which is outlined in \Fig{cfun23}(b). 

\InsertFig{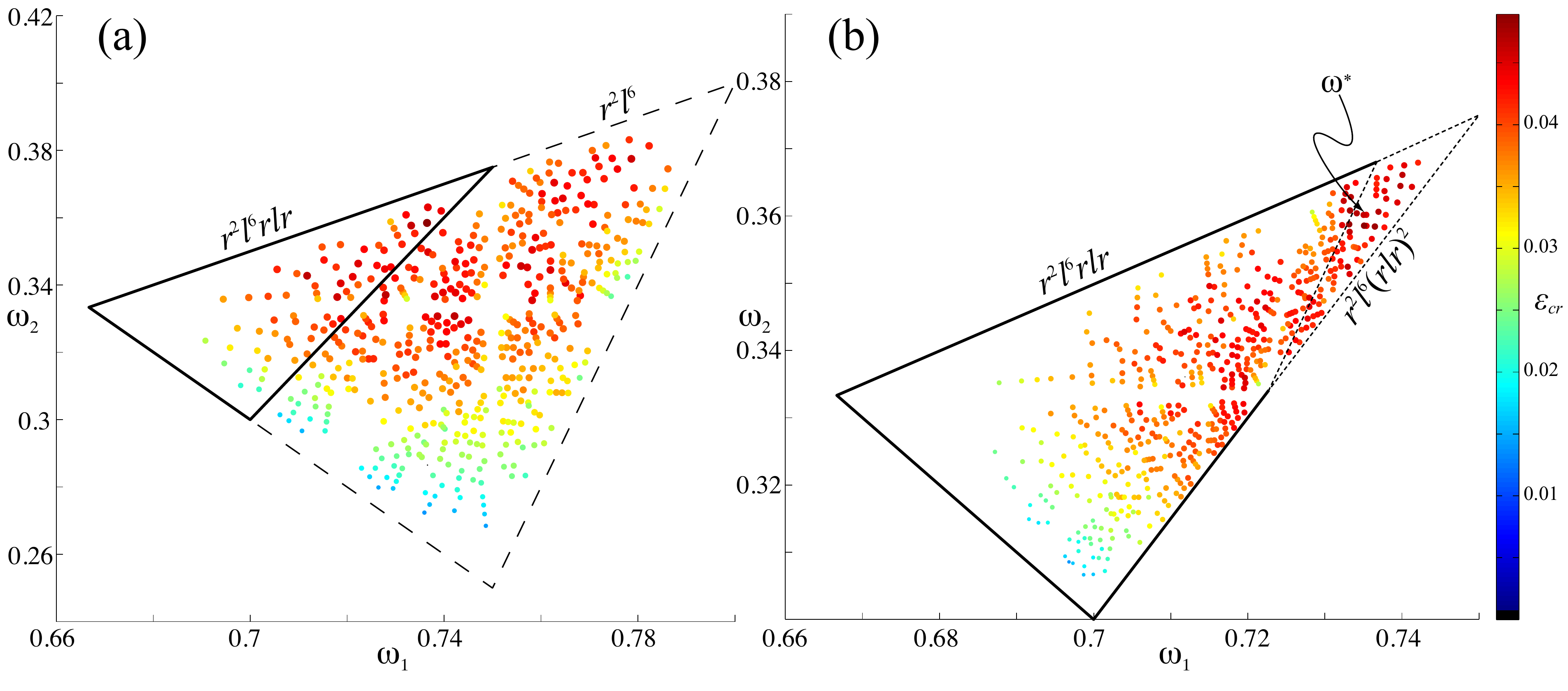}{Enlargements of the critical function for two ``most robust'' triangles, showing $\eps_{cr}$ for $512$ rotation vectors in the spiral field that have heads nine levels down from their respective triangles.
(a) Enlargement of the triangle $r^2l^6$ (dashed) from \Fig{cfun}. The level-$11$ triangle, $r^2l^6 rlr$ containing the most robust torus is also shown (solid). (b) Enlargement of the level-$11$ triangle from (a), and the level-$14$ triangle (small dashes) containing the most robust torus at this stage. The location of the most robust rotation vector at level-$38$ is denoted by $\omega^*$.}{cfun23}{\textwidth}

Continuing in this fashion, we increment the level of the base triangle by three and then find the most robust torus among $2^9$ rotation vectors within this triangle.
The peak of the critical function increases with each zoom, leading to the level-$29$ triangle
\beq{Level29}
	r^2l^6(rlr)^2 r (rl)^4 r^3 l^3
	\simeq \begin{pmatrix} 	172&241&330\\ 
							85&119&163\\ 
							234&328&449
	 \end{pmatrix} .
\eeq
The most robust torus in this triangle has path $p = r^2l^6 (rlr)^2 r (rl)^4 r^3l^3 r^2 (lr)^2 r^4$, giving  $(m,n) = (4800,2371,6531)$ so that $\omega \approx (0.7350,0.3630)$. This orbit has initial condition
\[
	(x^*,z^*,\delta^*) \approx
			(0, 0.123605791956645, -0.331972596409587).
\]
The peak critical value has converged to
$
	\eps_{max} = 0.0522 \pm 0.0005,
$
to three significant figures. Note that the given error bound signifies the observed variation in the maximal $\eps$ values for \emph{period $5000$ orbits}.  Since we fix the period, this $\eps_{max}$ does not necessarily represent the true maximal value of $\eps$ for infinite period. Nevertheless, this orbit is our best estimate, to period $5000$, of the most robust torus for \Eq{VPMap} for the standard parameters \Eq{StdParams}.  

To improve this result, we re-examined the level-$29$ triangle \Eq{Level29}. As before, we use the $2^9$ level $29+9$ periodic orbits within this triangle, but now add a tail $r^k$ so that the period is at least $25,000$.  Of these $512$ rotation numbers the most robust torus has the path 
\beq{LastTorusPath}
	p = r^2l^6 (rlr)^2 r (rl)^4 r^3l^3r^3(lr)^3 r^8,
\eeq
which agrees for the first $31$ levels with the previous one, but differs after that.
This corresponds to $(m,n)=(23749,11731,32316)$, so that
\[
	\omega \approx (0.73490, 0.36301).
\]
This torus is located at
\[
	(x^*,z^*,\delta^*) \approx
			(0, 0.123303207885153,-0.332181389896200)
\]
and is destroyed at the critical value. 
\beq{EpsCrMax}
	\eps_{max} = 0.0512 \pm 0.0005.
\eeq
This value of $\eps_{max}$ is less than the value of $\eps_{max}$ computed using the period 5000 orbits, which is consistent with the behavior seen in \Tbl{SpiralMeanEpsCR}. Namely, critical $\eps$ values, by and large, decrease as the period of the approximating periodic orbits increases, limiting on the true $\eps_{cr}$ from above.    

Unfortunately, unlike the noble numbers of the area-preserving case, there does not appear to be any simple pattern in the Farey sequence \Eq{LastTorusPath} of the most robust torus.

\subsection{Crossing Time}\label{sec:CrossingTime}
To compare these estimates for the peak values of $\eps_{cr}$ with the dynamics we performed a ``crossing time" experiment, similar to that of \cite{Meiss12a}. The point is that a rotational torus is a barrier so that any orbit that begins below the torus must remain below.

To make this computation easier, we choose a frequency map that is periodic mod-one in $z$, i.e., so that $\Omega(z+1) = \Omega(z) + m$ for some $m \in \bZ^2$. In this case, the invariant sets of the map in each unit interval of $z$ are identical. Moreover there are no rotational invariant tori whenever there exists an orbit for which $|z_t-z_0| > 1+\Delta$, where $\Delta$ is the maximal vertical extent of a torus. Since the first component of \Eq{Omega} satisfies the periodicity requirement, we modify only its second component:
\[
	\Omega_2(z) = \beta (\nint(z))^2-\delta,
\]
where ``$\nint$" is the nearest integer function. Since the tori for parameters near $(\eps,\delta) = (0.05,-0.3)$ are located near $z = 0.1$ and have vertical extent of order $0.1$, this modification will not affect their existence or nonexistence. Of course, it certainly changes the dynamics whenever $|z| > 0.5$.

The crossing time, $t_c$, is defined to be the first time for which $|z_{t_c}(x_0,z_0)-z_0| \ge 1.3$. If there is an orbit with $t_c < \infty$, the map has no rotational tori with $\Delta < 0.3$. The crossing time is certainly a highly variable function of initial conditions, so we compute its distribution for a set of initial conditions. Fixing $\delta = -0.33$, we start ten orbits with $x_0 = 0$ and a random value $|z_0| < 0.001$.  When $\eps \ge 0.1$ all of the orbits cross within a few thousand iterates, but as $\eps$ decreases, the crossing times grow rapidly, as shown in \Fig{CrossingTime}. We found that whenever $\eps < 0.0512$ none of the trajectories crossed within $10^{10}$ iterations. The implication, in complete agreement with \Eq{EpsCrMax}, is that there must be rotational tori in this case. The median of the distribution of $t_c$ for $0.0512 \le \eps \le 0.10$ can be fit to the power law
\[
	t_c = 0.0132(\eps-\eps_\infty)^{-3.807},
\]
with $\eps_\infty = 0.05047 \pm 0.0002$ as shown in \Fig{CrossingTime}. The value $\eps_\infty$ is close to \Eq{EpsCrMax}, and the difference is probably due to the weight given in the fit to points for larger values of $\eps$, though it may also be due to the sharp dependence of $\eps_{cr}$ on $\delta$.

Indeed, we compare the crossing time experiments for various values of $\delta$ with the critical $\eps$ values of the tori in \Fig{CrossingTime}(b). The points in the figure represent the computed $(\eps_{cr},\delta_{cr})$ for the 2048 tori of \Fig{cfun} in the $rrl$ and $llr$ triangles. The upper envelope of these points is an estimate of the most robust torus for a fixed $\delta$. Since these points are not from the zoomed-in triangles, like those of \Fig{cfun23}, they underestimate the maximal $\eps_{cr}$. The crosses in \Fig{CrossingTime}(b) show the computed the positions of the pole, $\eps_\infty$, in the crossing time. These are obtained from power law fits to crossing times experiments with a maximal iteration time of $10^{10}$. Note that the maximal value of $\eps$ is a highly sensitive function of $\delta$.

The crossing time experiments confirm that the most robust torus of \Eq{VPMap} for the standard parameters \Eq{StdParams} is destroyed near \Eq{EpsCrMax}.

\InsertFigTwo{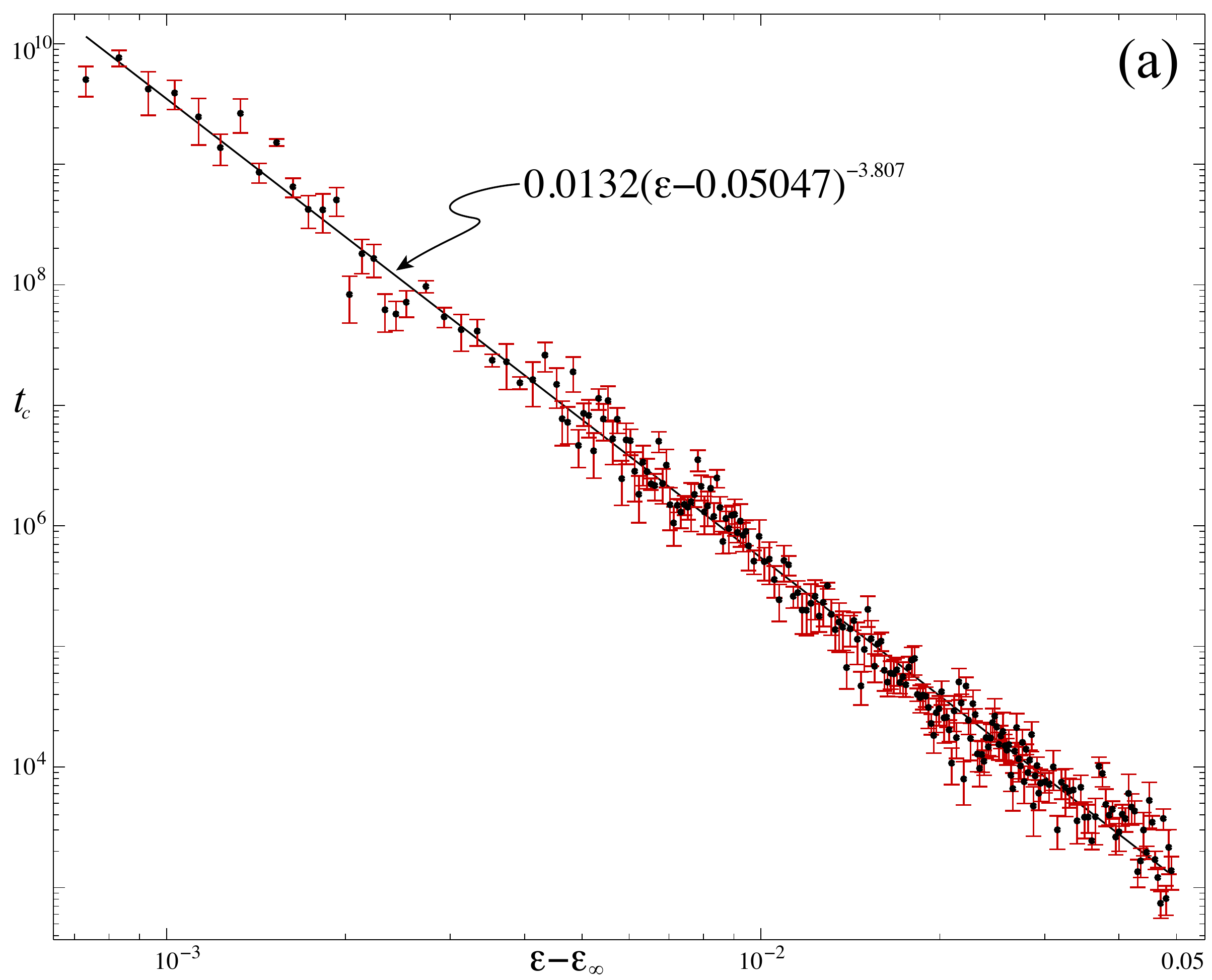}{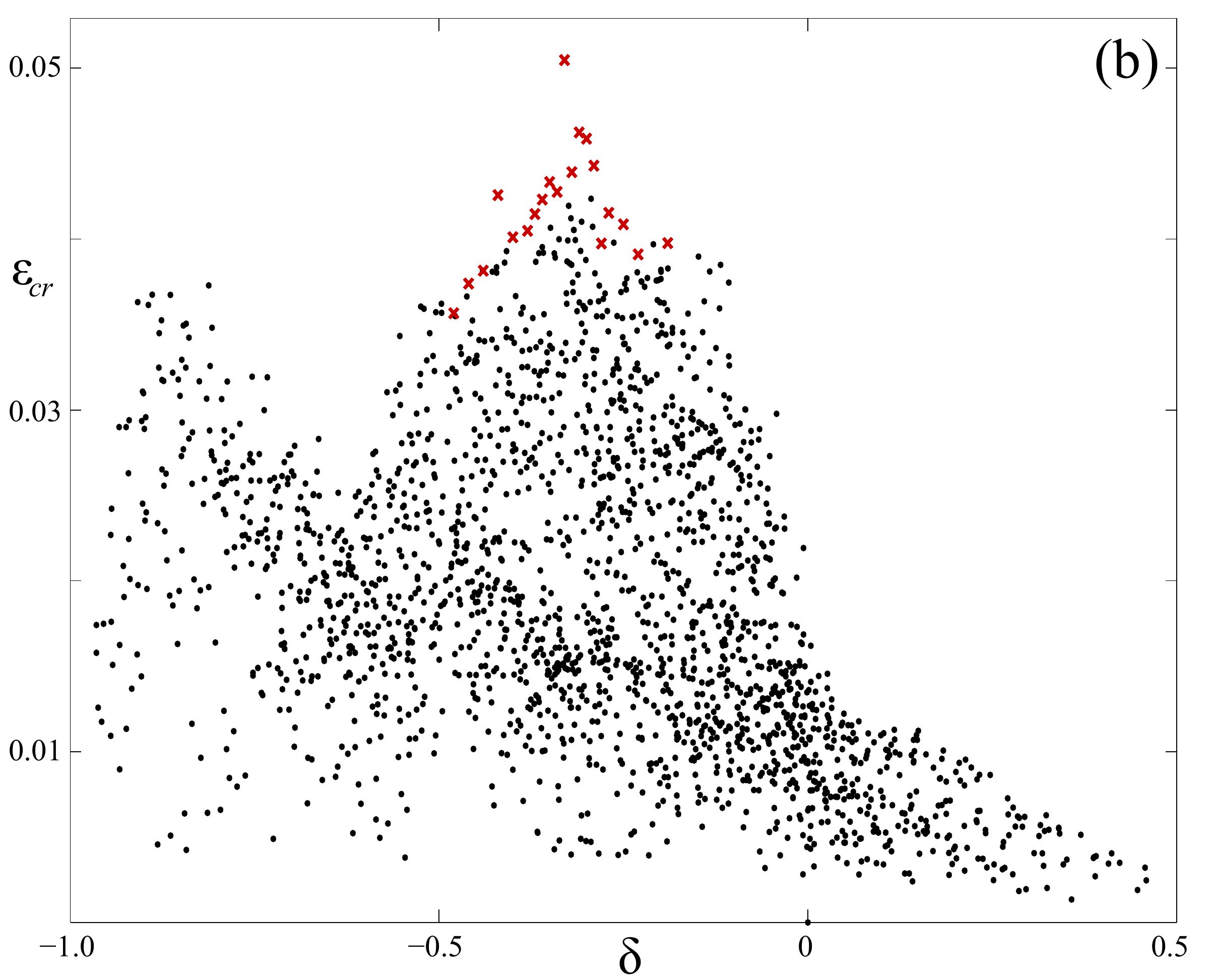}{(a) Crossing time as a function of $\eps$ for orbits of \Eq{VPMap} with $\delta = -0.33$. Points correspond to the median of the crossing times for each $\eps$, and error bars to the median absolute deviation. (b) Critical $\eps$ values as a function of $\delta$. The dots (black) are $(\eps_{cr},\delta_{cr})$ pairs for the tori of \Fig{cfun}, and the crosses (red) are estimates of the pole position, $\eps_{\infty}$, of $t_c$ for twenty values of $\delta$.}{CrossingTime}{3in}

\section{Conjectures and Questions}\label{sec:Conclusions}

Our investigation of the reversible, two-angle, one-action map \Eq{VPMap} has given support for several conjectures:

\begin{enumerate}
\item For each rational rotation vector $m/n$ in the range of $\Omega(z,\delta)$ and for each $\eps$, there is a $\delta$ such that \Eq{VPMap} has type a $(m,n)$-periodic orbit. This conjecture is the analogue of the Poincar\'e-Birkhoff theorem, and seems to be open for the volume-preserving case.

\item The residue \Eq{Residue} of each $(m,n)$ orbit of \Eq{VPMap} with force \Eq{Force} grows as $\eps$ or $\eps^2$ for $\eps \ll 1$ and as $\eps^{n-1}$ or $\eps^{n}$ for $\eps \gg 1$.

\item If the residue of a sequence of symmetric $(m,n)$-orbits limits to zero when $m/n \to \omega$ for a Diophantine rotation vector $\omega$, then there is a rotational invariant torus with that rotation vector. In other words, Greene's residue criterion holds. This conjecture is analogous to Tompaidis's theorem for the symplectic case \cite{Tompaidis96a}.

\item Conversely, if the residues of such a sequence are unbounded then the corresponding torus does not exist. This conjecture should be valid if the orbits converge to a remnant torus with positive Lyapunov exponent as shown by \cite{Falcolini92}.

\end{enumerate}

\noindent
There are also many questions:
\begin{itemize}
\item Is there an analogue of the dominant symmetry line for reversible maps like \Eq{VPMap}? More generally, is there a pattern to the signs of the residues and their oscillations for the various symmetric families of orbits?

\item Does the critical spiral torus have a particular threshold residue analogous to $R_{th} \approx 0.25$ for the noble tori of area-preserving maps?

\item Is there some significance to the oscillations in sign of the residues of the approximating periodic orbits, particularly near the cusps in the critical set shown in \Fig{ChangeC}?

\item Do the periodic orbits that are best approximants to a Diophantine rotation vector have any special significance in approximating the corresponding invariant torus?

\item Are there remnant tori analogous to the cantori of twist maps? The remnant torus shown in \Fig{Spiral1779} appears to be similar to a Sierpinski carpet. Is this true? It is known that symplectic maps of the form \Eq{NearIntegrable} have invariant Cantor sets when $\eps \ll 1$ for every rotation vector \cite{MacKay92b}, but no similar results are known for volume-preserving maps.

\item Is there a cubic field that, like the golden mean for twist maps, gives rise to locally most robust tori in the volume-preserving case? Several other fields, in addition to the spiral field that we have studied, have been proposed \cite{Lochak92, Tompaidis96b, Celletti04}.

\end{itemize}

\appendix
\newpage
\appendixpage

\section{Reversors} \label{app:appendix}
Suppose that a map $f: M \to M$, with $M = \bT^d \times \bR^k$ is reversible with involutions $S_1$ and $S_2$, and let $F$ be a lift of $f$ to $\bR^d \times \bR^k$.

Note that if $S$ is a reversor for $f$, then so is $S \circ f^t$ for any $t \in \bZ$; indeed this follows because $f^t$ is a symmetry of $f$. For the lift $F$, the translations \Eq{Rigid} are also symmetries; consequently, the translation of a reversor $S$ is also a reversor, $S \circ T_m$. Note that $S_i \circ T_m = T_{-m} \circ S_i$. These transformations generate a group
\[
	\langle S_1, S_2, F, T_m \rangle
\]
that is contained in the reversing symmetry group of $F$ whenever $g$ is odd.
For \Eq{VPMap} with \Eq{Omega} and \Eq{Force}, the reversors were given in \Eq{Involution}, This appears, as far as we know, to be the complete group.

The fixed set of $S$ is
\beq{FixSet}
	\fix{S} = \{(x,z): S(x,z) = (x,z)\}
\eeq
As is well-known, every symmetric periodic orbit has points on the fixed sets.
\begin{lem}\label{lem:FixedSets}
If $f$ has reversor $S$, then every symmetric orbit must have a point on either $\fix{S}$ or $\fix{f \circ S}$.
\end{lem}
\begin{proof}
 Indeed, if $j=2k$ is even then 
\[
	(x_k,z_k) = f^{-k} (x_{2k},z_{2k}) = f^{-k} \circ S (x_0,z_0) = S (x_k,z_k) ,
\]
so that $(x_k,z_k) \in \fix{S}$. If, on the other hand, $j = 2k-1$ is odd, then
\[
	 (x_k,z_k) = f^{-k+1} \circ S (x_0,z_0) = f \circ S (x_k,z_k),
\]
so that $(x_k,z_k) \in \fix{f \circ S}$.
\end{proof}

Even better, periodic orbits have points on two distinct symmetry curves separated by half of the period. 

\begin{lem}\label{lem:TwoFixedSets}
If $f: \bT^d \times \bR^k$ has a reversor $S$ and $\Gamma$ is a symmetric, $(m,n)$-periodic orbit of the lift $F$ to the universal cover $\bR^d \times \bR^k$, then $\Gamma$ has points on two fixed sets of reversors in the group generated by $\langle S, F, T_m \rangle$.
\end{lem}
\begin{proof}
Since $\Gamma$ is symmetric, we can suppose that there is a point $(x_0,z_0) \in \Gamma \cap \fix{S}$. When the period $n = 2\ell$ is even then, since $(x_{-\ell},z_{-\ell}) = T_{-m}(x_\ell, z_\ell)$,
\[
	(x_\ell,z_\ell) = F^\ell(S(x_0,z_0)) = S(x_{-\ell},z_{-\ell}) = S\circ T_{-m}(x_\ell,z_\ell) ;
\]
thus $(x_\ell,z_\ell) \in \fix{S \circ T_{-m}}$. Similarly when $n=2\ell-1$ is odd, then since
$(x_{-\ell+1},z_{-\ell+1}) = T_{-m}(x_{\ell},z_{\ell})$,
\[
	(x_{\ell},z_{\ell}) = F^{\ell}(S(x_0,z_0)) = F \circ S(x_{-\ell+1},z_{-\ell+1}) 
			= F \circ S\circ T_{-m}(x_{\ell},z_{\ell}) ;
\]
thus $(x_{\ell},z_{\ell}) \in \fix{F \circ S \circ T_{-m}}$.
\end{proof}

If we denote the symmetry by $S_1$ and let $S_2 = f \circ S_1$, then, the lemmas imply that symmetric orbits can be found by looking for orbits that start on $\fix{S_1}$ or $\fix{S_2} \equiv \fix{f \circ S_1}$, and ``half" a period later end on one of the sets $\fix{S_1 \circ T_{-m}}$ or $\fix{S_2 \circ T_{-m}}$. The results are summarized in \Tbl{Symmetry}.

\bibliographystyle{alpha}
\bibliography{breakup}{}

\end{document}